\documentclass[12pt]{article}
\RequirePackage[OT1]{fontenc}
\RequirePackage{amsthm,amsmath}
\RequirePackage{natbib}
\RequirePackage[colorlinks,citecolor=blue,urlcolor=blue]{hyperref}
\hypersetup{colorlinks = false}
\usepackage{graphicx}
\usepackage{booktabs}
\usepackage{fullpage}
\usepackage{pdflscape}
\usepackage{multirow}
\usepackage{longtable}
\usepackage{float}
\newcommand{\G}{\mathbf{G}}

\newcommand{\X}{\mathbf{X}}

\newcommand{\Hb}{\mathbf{H}}
\newcommand{\de}{\boldsymbol{\delta}}
\newcommand{\ta}{(1) Mend.$^{\text{a}}$}
\newcommand{\tb}{(2) GB$^{\text{b}}$}
\newcommand{\tc}{(3) GB with Mend.$^{\text{c}}$}
\newcommand{\td}{(4) Mend.$^{\text{d}}$}
\newcommand{\te}{(5) GB$^{\text{e}}$}
\newcommand{\tf}{(6) GB with Mend.$^{\text{f}}$}
\newcommand{\tg}{(7) Mend. without GC$^{\text{g}}$}
\newcommand{\thh}{(8) Mend. with GC$^{\text{h}}$}
\newcommand{\ti}{(9) Platt scaling}
\newcommand{\tj}{(10) Isotonic regression}
\DeclareMathOperator*{\argmin}{arg\,min}

\newcommand{\beginsupplement}{%
        \setcounter{table}{0}
        \renewcommand{\thetable}{S\arabic{table}}%
        \setcounter{figure}{0}
        \renewcommand{\thefigure}{S\arabic{figure}}%
        \setcounter{section}{0}
        \renewcommand{\thesection}{S\arabic{section}}%
}

\begin{document}

\title{Extending Models Via Gradient Boosting: \\ An Application to Mendelian Models}
\author{Theodore Huang$^{1,2,*}$, Gregory Idos$^{3}$, Christine Hong$^3$, \\ Stephen Gruber$^3$, Giovanni Parmigiani$^{1,2}$, Danielle Braun$^{1,2}$}
\date{}

\maketitle

\begin{center}
$^{1}$Department of Biostatistics, Harvard T.H. Chan School of Public Health \\
$^{2}$Department of Data Science, Dana-Farber Cancer Institute \\
$^3$City of Hope Comprehensive Cancer Center \\
$^*$Corresponding author: thuang@ds.dfci.harvard.edu
\end{center}

\begin{abstract}
Improving existing widely-adopted prediction models is often a more efficient and robust way towards progress than training new models from scratch.
Existing models may (a) incorporate complex mechanistic knowledge, (b) leverage proprietary information and, (c) have surmounted barriers to adoption.
Compared to model training, model improvement and modification receive little attention. 
In this paper we propose a general approach to model improvement: we combine gradient boosting with any previously developed model to improve model performance while retaining important existing characteristics. 
To exemplify, we consider the context of Mendelian models, which estimate the probability of carrying genetic mutations that confer susceptibility to disease by using family pedigrees and health histories of family members.
Via simulations we show that integration of gradient boosting with an existing Mendelian model can produce an improved model that outperforms both that model and the model built using gradient boosting alone.
We illustrate the approach on genetic testing data from the USC-Stanford Cancer Genetics Hereditary Cancer Panel (HCP) study.
\end{abstract}

\section{Introduction} \label{intro}

Classification and prediction models are playing increasingly important roles in science and in a range of applications across society. With the expanding availability of data, statistical and machine learning theory have emphasized training new prediction models. The large number of models in practical use, however, creates a need for techniques that can improve them without rebuilding them, for example by revising their structure or adding features. Existing models may already embed complex mechanistic relationships between variables or may have been extensively trained using large amounts of data, proprietary data, or data that require extensive curation. Training a new model from scratch may fail to take advantage of valuable previous work; accordingly, building a new model that outperforms the existing model may be challenging. At the same time, improvements and adaptations are often needed. Two typical examples in biomedicine are the emergence of new biomarkers and the need to adapt a model for applications to slightly different populations \citep{doi:10.1093/biostatistics/kxy044}.
Using existing models can also be advantageous from an implementation perspective. For example, clinicians rely on prediction models to better understand patients' risks of developing diseases and to accordingly make informed decisions. Adoption of these models for clinical use is a laborious process. A new model may have to surmount logistical, reputational and sometimes commercial obstacles before being adopted for clinical use, while clinicians may more quickly and readily adopt a modification or upgrade of a trusted existing model.

Although directly improving prediction models is crucial, in practice it may not be straightforward. Incorporating new data or new features to existing models often requires the original training data, and obtaining these data can be impractical. In addition, some models may be complex and depend on prior scientific evidence. Consequently, incorporating new features in the same manner that the current features are utilized in the existing model may be challenging due to limitations in scientific knowledge. Lastly, the modeling mechanisms of some models are  proprietary, and hence improving these models by modifying their structure can be infeasible.

Several approaches have been proposed to address this issue. \cite{su2018review} and \cite{janssen2008updating} both analyze a variety of updating methods for clinical prediction models. These methods can update an existing model to be more applicable on a new population and incorporate new features; however, these approaches are limited to regression models and would not necessarily apply to black box models. Methods to calibrate existing black box prediction models have also been developed, such as Platt scaling \citep{platt1999probabilistic} and isotonic regression \citep{ayer1955empirical,brunk1955maximum}. Although these methods can improve black box models, they cannot incorporate new features.

With these challenges in mind, in this paper we explore the idea of developing a machine learning technique for model improvement of an arbitrary given model--the reference model. The reference model could be a black box, as long as predictions can be computed at every point in the feature space. 
The general structure of our proposal is to start with a previously trained and successful model and a new training data set for the desired improvement. For example, the new data set could include additional information on a newly discovered biomarker but consist of far fewer observations than the original data used to train the model. We then compute predictions from the reference model and their residuals, and fit the residuals using a simpler technique that allows for easy adjustment of the predictions. 
In this paper we provide a pilot implementation of this concept using Gradient Boosting (GB) \citep{friedman2000additive}, a popular ensemble prediction algorithm for combining simple individual component models, or ``weak learners", to produce a more accurate prediction model. The algorithm iteratively identifies a current model's weaknesses and learns how additional data can be leveraged to incrementally overcome these weaknesses, which makes it ideal for our task. The algorithm is typically used to build prediction models from scratch; as far as we know, it has not been applied to improving existing models.

The application that motivated this development is predicting genetic predisposition to cancer. Some cancers are caused by inherited germline mutations, and thus a family history of cancer can inform individuals of their risk of carrying a mutation in a cancer susceptibility gene. Cancer geneticists and genetic counselors help individuals with a family history of certain cancers better understand their risk of having a genetic predisposition to these cancers and formulate a plan for genetic testing, screening and prevention. Statistical models that can accurately and expeditiously predict this risk are thus critical tools for health care providers to advise counselees.

Numerous risk prediction models have been developed to predict genetic predisposition to cancer \citep{Kastrinos2018,parmigiani2007validity}. Some of these models directly use training data on genetically tested individuals to estimate empirically the probability of carrying a mutation given the cancer history, usually through regression models \citep{couch1997brca1,vahteristo2001probability,barnetson2006identification}. Other models are Mendelian and use the age-specific probability of developing the cancer given the genotype, called penetrance; the population-level distribution of the genotypes, called prevalence; and Mendelian laws of inheritance to estimate the carrier probability \citep{murphy1969application}. Examples of Mendelian models include BOADICEA \citep{antoniou2008boadicea} and {MMRpro} \citep{chen2006prediction}.

Mendelian models are complex and embed the pedigree structure to acknowledge the inherited pattern of the cancer susceptibility mutations. As a result, it can be challenging to directly improve such models by changing the way they account for existing features, or by accounting for additional features. GB provides a means for empirically incorporating additional information to directly improve upon the existing Mendelian model predictions. The idea is to initialize the GB algorithm with the predictions from the existing Mendelian model, and then train the new model to learn how the information from the features can correct the shortcomings of the original model. In this work, we explore the potential symbiotic relationship between GB and Mendelian prediction--how GB can improve Mendelian prediction, and how Mendelian prediction can improve GB. The models are introduced in Sections \ref{gb} and \ref{mendel}. In Section~\ref{sim}, we conduct a simulation study to compare the effectiveness of GB and Mendelian models in incorporating cancer family history information. In Section~\ref{data}, we apply this approach to data from the USC-Stanford Cancer Genetics Hereditary Cancer Panel (HCP)  study \citep{idos2018promoting}. Finally, we conclude with a discussion in Section~\ref{discussion}.

\section{Gradient Boosting} \label{gb}

Gradient boosting (GB) is an iterative ensemble learning method that combines boosting, which is a machine learning technique that sequentially adds weak learners to create a strong learner, and gradient descent, which is an iterative optimization procedure that takes steps proportional to the negative gradient to find local minima. Consider training data $(z_i, y_i)$ for $i=1, \dots, N$, where $y_i$ is the outcome and $z_i$ is a feature or vector of features. The goal is to obtain a prediction $P(z_i)$ of $y_i$. GB first initializes the prediction as $P_0(z_i)$. Then at the $m$-th iteration, it calculates the residuals $r_{im} = -\left[ \partial L(y_i, P(z_i)) / \partial P(z_i) \right]_{P(z_i) = P_{m-1}(z_i)}$ between the outcome $y_i$ and the prediction $P(z_i)$, where $P_{m-1}(z_i)$ is the prediction at the $(m-1)$-th iteration, based on a specified loss function $L$. GB then fits a base learner $h_m(z_i)$ (often a decision tree) using $(z_i, r_{im})$, $i = 1, \dots, N$, learning the residuals through the features. Lastly, the new learner is added to the current prediction via $P_m(z_i) = P_{m-1}(z_i) + \gamma_m h_m(z_i)$, where
\[\gamma_m = \argmin_{\gamma} \sum_{i=1}^N L(y_i, P_{m-1}(z_i) + \gamma h(z_i)).\]
This process is repeated, and after $M$ iterations, the final predictions are $P_M(z_i) = P_0(z_i) + \sum_{m=1}^M \gamma_m h_m(z_i)$.
This prediction model can then be applied to testing data. Overall, GB identifies the weaknesses of the current prediction model by using the residuals, and uses features in the data to predict these residuals and hence overcome the initial model weaknesses.

The choice of the loss function $L$ depends on the type of outcome variable $y$. For continuous outcomes, common choices of $L$ are squared-error loss, absolute loss, and Huber loss; for binary outcomes, common choices are logistic loss and Adaboost loss. \cite{natekin2013gradient} provide an overview of commonly-used loss functions in GB.

There are several options for fitting the base learners to the residuals. Linear regression is an example; however, a sum of linear regression models is still linear, and thus the boosting process of sequentially adding linear models would not lead to a more general class of models. An alternative choice is decision trees, which partition the feature space into rectangles by making linear splits. These are more commonly used in GB because they are more robust to overfitting, as their depth (i.e., the number of splits) can be controlled. Decision trees more naturally fit the goal of GB of combining weak learners to create a stronger one. \cite{natekin2013gradient} provide a complete discussion of the various options for the base learners.

In addition to the maximum decision tree depth or the number of iterations $M$, GB can also be customized through other tuning parameters. We can introduce a shrinkage parameter $\nu$ such that each $\gamma_m$ is changed to $\nu \gamma_m$, thereby decreasing the learning rate in each iteration. We can also sample a fraction $f$, called the bag fraction, of the training set at each iteration in the algorithm. Both of these tuning parameters are regularization methods aimed at reducing overfitting.

\sloppy Without a prior prediction, GB commonly initializes with $P_0(z_i) = \argmin_\gamma \sum_{i=1}^N L(y_i, \gamma)$. In the case of continuous outcomes with squared-error loss, this is the sample mean $\bar{y}$; in the case of binary outcomes with logistic loss, this is the sample log odds $\log(\bar{y} / (1 - \bar{y}))$. Alternatively, we propose here to use GB in combination with existing successful models by using these models to initialize the algorithm.

\section{Mendelian Risk Prediction Models} \label{mendel}

Mendelian risk prediction models are used to predict an individual's risk of carrying a mutation in a cancer susceptibility gene. Unlike empirical prediction models, they incorporate family history information by acknowledging the Mendelian inheritance pattern of genetic mutations in autosomal genes. In this section, we explain how Mendelian models predict risk and how existing Mendelian models can be used in combination with GB.

\subsection{Notation} \label{not}

Consider making predictions for an individual, the counselee, with $n$ family members. Let $\G_j$ denote the genotype for the $j$-th family member, where $j=1,\dots,n$. The counselee has index $j=1$. $\G_j$ can be a vector representing multiple genes, where each component of the vector is a binary indicator of carrying the corresponding mutation, or a ternary indicator of zygosity. The family's genotype configuration is  $\G = (\G_1, \dots, \G_n)$. Let $T_{rj}$ denote the time that the $j$-th individual develops cancer of the $r$-th type, $r = 1, \dots, R$. We consider discrete ages $1, 2, \dots, T_{max}$. Let $C_j$ denote the censoring time (either the time of death or the current age) for the $j$-th individual, and let $X_{rj} = \min(T_{rj}, C_j)$ be the observed time for the $j$-th individual for the $r$-th disease. Let $\X_j = (\X_{1j}, \dots, \X_{Rj})$ and $\X = (\X_1, \dots, \X_n)$. Let $\delta_{rj} = I(T_{rj} \leq C_j)$, $\de_j = (\de_{1j}, \dots, \de_{Rj})$, and $\de = (\de_1, \dots, \de_n)$. Let $\Hb_j = (\X_j, \de_j)$ be the observed age and cancer status for the $j$-th individual, and let $\Hb = (\Hb_1, \dots, \Hb_n)$ be the complete family history.

\subsection{Mendelian Carrier Probability Estimation} \label{cp}

Mendelian risk prediction models estimate the counselee's genotype probability conditional on their family history via
\begin{align} \label{eqn.cp}
\nonumber    P(\G_1 | \Hb) &= \frac{P(\Hb | \G_1) P(\G_1)}{\sum_{\G_1} P(\Hb | \G_1)P(\G_1)} \\
 \nonumber   &= \frac{\sum_{\G_2, \dots, \G_n} P(\Hb | \G) P(\G_2, \dots, \G_n | \G_1) P(\G_1)}{\sum_{\G_1} \sum_{\G_2, \dots, \G_n} P(\Hb | \G) P(\G_2, \dots, \G_n | \G_1) P(\G_1)} \\
    &= \frac{P(\G_1) \sum_{\G_2, \dots, \G_n} P(\G_2, \dots, \G_n | \G_1) \prod_{j=1}^n P(\Hb_j | \G_j) }{\sum_{\G_1} P(\G_1) \sum_{\G_2, \dots, \G_n} P(\G_2, \dots, \G_n | \G_1) \prod_{j=1}^n P(\Hb_j | \G_j)}
\end{align}
\citep{chen2004bayesmendel}. Here $\G$ is taken to be unknown, although extensions to partly known $\G$ are straightforward.
These models use Mendelian laws of inheritance to compute the relatives' genotype distribution conditional on the counselee's genotype: $P(\G_2, \dots, \G_n | \G_1)$. $P(\G_1)$ is the counselee's marginal (with respect to the family members) probability of carrying the genotype $\G_1$ and is taken to be the population prevalence. $P(\Hb_j | \G_j)$ is the age-specific probability of observing the cancer status and observed age given the genotype, which is obtained using the penetrance. Note that we make the fundamental assumption of conditional independence of the family members' family histories given the genotypes, which may not be true due to risk heterogeneity from genetic, environmental, or behavioral factors.
Both the prevalence and penetrance are usually derived from studies in the literature. Since the summation over all genotypic configurations in the family can be computationally intensive, Mendelian models often use the Elston-Stewart peeling algorithm \citep{elston1971general} to efficiently calculate the counselee's genotype distribution.

Mendelian models can incorporate information from any cancer and gene, as long as there are reliable corresponding penetrance estimates. In practice, for a given set of cancer susceptibility genes, there may only be one or two cancers that have reliable penetrance estimates ($R = 1$ or $2$). However, Mendelian models are flexible, and a user can easily include information from additional cancers by inputting their corresponding penetrances.

\subsection{Gradient Boosting with Mendelian Models} \label{gbmendel}

We propose to use GB to improve existing Mendelian models, for example by adding new features or overcoming limitations of restrictive structural assumptions. For each family, we define the outcome as the binary indicator $y$ of the counselee carrying at least one of the genetic mutations and $z$ are features that are associated with carrying the mutations (the outcome). These features can include ones already incorporated in the existing Mendelian model as well as new ones. For example, $z$ can contain family history information for cancers already considered in the existing Mendelian model as well as additional cancers not considered. 
Besides cancer information, $z$ can also include risk modifiers such as medical interventions that impact risk.

We propose to initialize the GB algorithm with the existing Mendelian model predictions $P(\G_1 | \Hb)$ from equation \ref{eqn.cp}. Here $\Hb$ is the family history information for only the cancers used in the Mendelian model. The initial prediction in GB is typically not a function of the features $z$ (as seen in Section~\ref{gb}); in contrast, as we use GB to improve an existing model, we set the initial prediction to the Mendelian prediction, a function of the input~$\Hb$. For the base learner $h_m$, we demonstrate our method using decision trees with features $z$. Since our outcome is binary, we use the logistic loss function $L(y, \theta) = \log(1 + \exp(\theta)) - y\theta$,
where $\theta$ is the log odds of being a mutation carrier. The logistic loss function is the negative of the likelihood for the Bernoulli distribution, and the residual is $-\partial L(y, \theta) / {\partial \theta}  = y - [ 1 + \exp(-\theta) ]^{-1}$.

\section{Simulation Study} \label{sim}

\subsection{Gradient Boosting for Incorporating Family History Information} \label{addcan}

To assess the performance of our approach, we conduct a simulation study, with a focus on exploring various methods of incorporating cancer family history information to predict the mutation carrier status. Mendelian models rely on penetrances for cancers related to the genetic mutations of interest. These are often estimated from published studies  \citep{chen_meta-analysis_2007,marroni2004penetrances}, which can be challenging due to low mutation prevalences, difficulty in finding data cohorts with genetically tested individuals, bias in cohort ascertainment, and heterogeneity in study populations. Mendelian models should ideally only incorporate information for cancers with accurate and precise penetrance estimates, as using inaccurate penetrances can lead to poor model performance. Applying stringent criteria on the quality of input penetrances may lead to the omission of potentially useful information about cancers known to be associated with a genetic mutation, but for which there is not enough evidence in the literature to estimate the penetrance. Using inaccurate penetrances estimates when adding cancers to a model could fail to improve predictive performance or even worsen it.

In settings where the cancer penetrances are potentially inaccurate, GB can be an effective alternative approach to incorporating family history information. GB does not need to rely on complete age-specific penetrance estimates but instead can integrate the information in an empirical manner. It can thus overcome two main limitations of existing Mendelian models: (1) potential inaccuracies of penetrances of the cancers already incorporated in the model, and (2) difficulty in including information from additional cancers. Using the notation in Section~\ref{gbmendel}, we let the features $z$ represent the family history information from all the cancers of interest. These cancers can include both the ones used in the existing Mendelian model as well as others for which we may have family history information but no accurate penetrance estimates. Suppose the existing Mendelian model uses information from $R$ cancers, and we are also interested in incorporating information from an additional $R'$ cancers (new features). One approach to incorporating the entire cancer family history information is by defining $z = (z_{1}, \dots, z_{R}, z_{R+1}, \dots, z_{R+R'})$, where $z_{r}$ is the proportion of family members who develop the $r$-th cancer (for sex-specific cancers, we only consider family members of that sex), $r = 1, \dots, R + R'$. Here $z_{1}, \dots, z_{R}$ represents cancers in the existing Mendelian model and $z_{R+1}, \dots, z_{R+R'}$ represent the additional cancers (new features). In addition or in alternative to these proportions, other approaches to incorporating family history information into the features $z$ can also be used.

We use simulations to compare the GB and Mendelian approaches to incorporating additional cancers. We also analyze the combination of GB and Mendelian models, where the GB algorithm is initialized with the Mendelian predictions, and explore settings in which the added GB component can overcome deficiencies in the existing Mendelian model. In addition, we compare the effectiveness of the GB approach in adding new cancers with the Mendelian approach of adding these cancers through penetrance estimates.

We evaluate the proposed approach using MMRpro, a Mendelian model which predicts mutation carrier status of the \textit{MLH1}, \textit{MSH2}, and \textit{MSH6} genes \citep{chen2006prediction}. These mutations are related to DNA mismatch repair (MMR) and have been shown to be linked with Lynch syndrome \citep{papadopoulos1994mutation,fishel1993human,miyaki1997germline}, a hereditary cancer syndrome that predisposes individuals to increased risks of many cancers, most commonly colorectal (CRC) and endometrial (EC), but also gastric, ovarian, small intestine, and others \citep{lynch1996hereditary}. {MMRpro} uses family history information on CRC and EC ($R = 2$), as these cancers have been extensively studied and hence have reliable penetrance estimates derived from a meta-analysis \citep{chen2006prediction}. However, {MMRpro} ignores family history information on other Lynch syndrome cancers whose penetrance estimates are not as reliable. We focus on gastric cancer (GC), which has not been studied as extensively as CRC and EC, although penetrance estimates \citep{braun2018clinical,dowty2013cancer,barrow2009cumulative} have been published. In this work, we incorporate family history information on GC ($R' = 1$) to the existing MMRpro predictions (using version 2.1-5 of the BayesMendel R package), both through GB and through a completely Mendelian approach.

\subsection{Generating Families\label{genfam}}

We generate family data sets with two sample sizes: 10,000 and 1,000. Each family consists of 3 or 4 generations. The core consists of an index case, or counselee, mother, father, and maternal and paternal grandparents. We then randomly sample the number of sisters, brothers, maternal and paternal aunts and uncles, daughters, sons, and each sibling's number of daughters and sons from $\{0, 1, 2, 3\}$. Thus, family sizes range from 7 to 67 and reasonably mimic the variability seen in real data.

We first generate genotypes for the family members. For these simulations, we use inflated allele frequencies of 0.01 for each of the three MMR mutations in order to increase the number of carriers and thus enable us to better address our methodological aims. After having generated genotypes for the founders of the pedigree, we generate the genotypes for the rest of the family using the Mendelian laws of inheritance.

Using these genotypes, we then generate cancer status and age for every individual in the family for CRC, EC, and GC. For CRC and EC, we generate data using two scenarios: one where the cancers have high penetrance, to mimic situations where the cancer information is strongly predictive of the mutation carrier status, and another where the cancers have low penetrance to produce a small amount of signal for prediction, leaving potential for improved performance when incorporating additional cancers. To simulate high-penetrance cancers, we use the MMRpro penetrances, and to simulate low-penetrance cancers, we use scaled versions of the MMRpro penetrances, where the lifetime risk for each of the three genotypes corresponding to carrying exactly one of the three MMR mutations is 0.2.
For GC, we base the data-generating penetrances on estimates from \cite{dowty2013cancer,moller2018cancer} for carriers. For non-carriers, we use penetrances from the Surveillance, Epidemiology, and End Results (SEER) program \citep{devcan,fay2003age,fay2004estimating}. We scale the GC penetrances so that the lifetime risk of non-carriers is 0.05 and the lifetime risk for each of the three genotypes corresponding to carrying exactly one MMR mutation is 0.5. We use this scaled version of the penetrance in order to amplify the association between carrying an MMR mutation and developing GC. This helps us clearly evaluate the ability of GB to leverage information from the additional GC. After obtaining the penetrances for non-carriers and carriers of exactly one of the MMR mutations, we obtain penetrances for carriers of multiple mutations by multiplying the survival functions of the individual mutations, thus calculating penetrances for all $3^3 = 27$ genotypes.

Using these penetrances, which are defined from ages 1 to 94, we generate cancer ages for the family members (95 if they do not develop the cancer). We also generate current ages for all individuals. The counselee's grandmothers (who are founders in our pedigrees) have their current ages generated from a truncated normal distribution with mean 100 and variance 4, and current ages of their spouses (as well as ages of other spouses in the pedigree) are generated from a normal distribution with the spouse's current age as the mean and variance 4. All children are generated iteratively using a normal distribution with the mother's current age minus 30 as the mean and variance 25 (or simply assigned to be the mother's current age minus 15, if this is less than the generated current age).
Supplementary Table~\ref*{stab:simfams} summarizes features of the simulated families for both the low and high-penetrance scenarios.

\subsection{Simulation Design} \label{simsetup}

We generate family data using the above approach and then randomly select half of the data to be used for training (for the GB models) and the other half to be used for testing all models.
For each simulated counselee we estimate the mutation carrier probability through a variety of models, using both GB and Mendelian approaches. The Mendelian models are split into two scenarios: the first uses misspecified (different from data-generating) penetrances for the existing cancers (CRC and EC), obtained by taking the square root of the survival functions and converting back to penetrances (see Supplementary Section~\ref*{ssec.mis} for details), and the second consists of ``oracle" (same as data-generating) Mendelian models with the correctly specified penetrances. Using misspecified penetrances allows us to compare the GB approach to incorporating family history information to the Mendelian approach when their  predictive ability is reduced. It also allows for assessing the performance of GB when revising the original model on a population different from that used to develop the model. A list of all prediction models used in the simulation analysis is in Supplementary Table~\ref*{stab:mod}, with eight main models described below. \\

\noindent \textit{Models Without GC, and With Misspecified CRC and EC Penetrances} \\
We first explore models that only incorporate CRC and EC and ignore GC. These include: (1) the Mendelian model with misspecified CRC and EC penetrances, without using GC information; (2) GB without the Mendelian predictions (initialized with the sample log odds), using information from only CRC and EC as features; and (3) GB initialized with the Mendelian predictions in model~1, using information from only CRC and EC as features. \\

\noindent \textit{Models With GC Information, Misspecified CRC and EC Penetrances} \\
Next, we explore models that incorporate all three cancers, including GC. Thus we analyze the ability of the models to integrate information from additional cancers. This category contains three models: (4) the Mendelian model with misspecified CRC and EC cancer penetrances, incorporating GC information through the data-generating penetrance; (5) GB without the Mendelian predictions, using information from all three cancers as features; and (6) GB initialized with the Mendelian predictions in model~1, using information from all three cancers as features. In addition to these three models, we also conduct a sensitivity analysis examining situations when the GC penetrance is misspecified in the Mendelian model. The GC penetrances are misspecified by raising the survival functions to various exponents and then converting back to penetrances.
Supplementary Figure~\ref*{sfig_pen_gc} visualizes the misspecified penetrances. \\

\noindent \textit{Oracle Mendelian Models, Correctly-specified CRC and EC Penetrances} \\
We also compare the previous models with ``oracle" Mendelian models where the penetrances are all correctly specified. These models represent the ideal case and are a gold standard with which we can compare the other approaches. We consider two oracle models: (7) the Mendelian model incorporating only CRC and EC through the data-generating penetrances; and (8) the Mendelian model incorporating all three cancers through the data-generating penetrances. \\

\noindent \textit{Other updating methods} \\
Lastly, we compare all the models with (9) Platt scaling and (10) isotonic regression, two commonly used methods to calibrate prediction models. We use these methods to update model 1. \\

We consider two additional simulation scenarios. In one, we generate the training sets with the high penetrances and the testing sets with the low penetrances (as described in Section~\ref{genfam}). This is to assess GB's predictive performance when the updated model is tested on a population that is different than that in which it was trained. In the other scenario, we generate gastric cancer using a low penetrance and use a smaller sample size of 1,000 families. We use the SEER GC penetrance for noncarriers \citep{devcan,fay2003age,fay2004estimating}, and multiply the noncarrier penetrance by 2 to obtain the carrier penetrances. This investigates the situation where the training data have limited information about the new features. This is the case in the data application in Section~\ref{data}, where we have few individuals with gastric cancer (see Table~\ref{tab_usc}). Thus, this simulation attempts to mimic the data application and explain some of the results from Section~\ref{results}.

To run GB, we use XGBoost \citep{chen2016xgboost}, a scalable and popular implementation of GB, through the {\tt xgboost} R package.  For the base learners, we use decision trees with three candidate features: the proportion of family members with CRC, the proportion of female family members with EC, and the proportion of family members with GC. We limit each tree to a maximum depth of 2, the minimum depth allowing for interactions between features. We also set the shrinkage parameter to $\nu = 0.1$ and the bagging fraction to $f = 0.5$. These choices are commonly used and based on suggestions in \cite{friedman2001greedy}, \cite{friedman2002stochastic}, and \cite{friedman2001elements}. The main GB models described above use 50 iterations. We provide a sensitivity analysis for the number of iterations in the Supplementary Materials. To account for the variability in the selection of the training set, we use 100 Monte Carlo cross-validation replicates for the entire algorithm. In each replicate we randomly select half of the data to be the training data and the other half to be the testing data. We then average the performance metrics over these 100 replicates. We also compare this with a bootstrap approach \citep{steyerberg2001internal}. Here the estimated performance is the difference between the apparent (trained on whole data, tested on whole data) performance and the average difference of the bootstrap (trained on bootstrap sample, tested on bootstrap sample) and test (trained on bootstrap sample, tested on whole data) performances. Lastly, to assess the uncertainty of our performance measures, we generate 100 data sets to obtain percentile confidence intervals for each performance measure.

\subsection{Simulation Results\label{simres}}

\subsubsection{Performance Measures} \label{perfmeas}

We evaluate model performance on the testing data using metrics to assess calibration, discrimination, and overall performance \citep{steyerberg2010assessing}. To rigorously assess calibration, we use metrics that vary on the level of strictness: we use the ratio of the number of observed to expected events (O/E) to measure mean calibration (``calibration-in-the-large"), calibration intercepts and slopes to measure weak calibration, and calibration plots (split into risk deciles) to measure moderate calibration \citep{van2016calibration}. The mean and weak calibration metrics will lead to correct calibration if the training and testing sets are similar, and hence assessing moderate calibration is necessary. We assess discrimination using the area under the ROC curve (AUC) and assess overall performance using the root Brier score (rBS), defined as $\sqrt{\sum_i (y_i - P_M(z_i))^2}$.

\subsubsection{Main Analysis} \label{simres_main}

Table~\ref{tab_sim} summarizes results of our simulation for models 1 to 10, as described in Section~\ref{simsetup}, in two data-generating scenarios for CRC and EC. It provides average performance measures across the 100 simulated data sets, along with 95\% confidence intervals based on percentiles. For each model and metric, we also include the percentage of simulated data sets in which the model outperformed the baseline Mendelian model without GC information (model 1).
Since we used misspecified CRC and EC penetrances, model 1 has relatively poor performance. GB, without the Mendelian prediction and without GC information (model 2), has a better O/E but worse AUC and rBS. The combination of GB and the Mendelian model, using only CRC and EC as features (model 3), combines the strengths of the two approaches by having the best O/E while matching the AUC and rBS of the Mendelian model (model~1).

We can also assess the ability of the two approaches to incorporate GC information. The completely Mendelian method of using a GC penetrance (model 4) (where we use the true data-generating GC penetrance) shows improvement in O/E, AUC, and rBS over the Mendelian model without GC information (model 1). Again, the GB approach without the Mendelian prediction, incorporating all three cancers as features (model 5), had a better O/E but worse AUC and rBS compared to model 4. The GB and Mendelian model combination (model 6) again produced the best of the three models with GC information, with an improved O/E and similar AUC and rBS values compared to the completely Mendelian model that incorporates GC information (model 4). Platt scaling and isotonic regression did not incorporate gastric cancer information; thus, although they performed well in O/E, they did not improve in AUC.
These results illustrate how misspecified penetrance inputs which are systematically lower than the correct data-generating model can affect the mean calibration of Mendelian models.

\begin{figure}
    \label{sfig:cal}
    \centering
    \includegraphics[width=\linewidth]{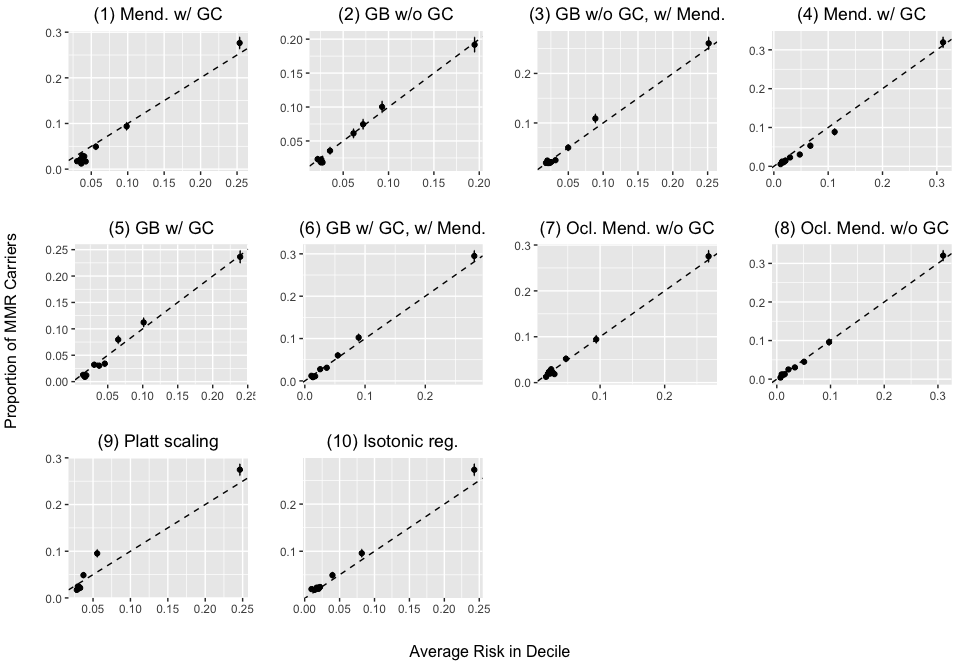}
    \caption{Calibration plots for the simulated data with low-penetrance data-generating CRC and EC, for 10 Monte Carlo cross-validation replicates. Each point represents a risk decile, and the error bars are the binomial confidence intervals for the observed proportion of carriers. Each family may appear multiple times in the plot if they were in the testing set multiple times. The results are for a single simulated data set.}
\end{figure}

We further assess calibration using calibration intercepts, slopes, and plots \citep{van2016calibration} in Supplementary Table~\ref*{stab:cal} and Supplementary Figure~\ref*{sfig:cal}. Again, we see that the GB models have better intercepts and slopes than the Mendelian models with misspecified CRC and EC penetrances (models 1 and 4). Platt scaling (model 9), which does not use any gastric cancer information, also performs well in both metrics, as do the oracle Mendelian models.

When looking at the calibration plots, we see that the Mendelian models with misspecified CRC and EC penetrances (models 1 and 4) consistently underpredict across the range of risks. On the other hand, GB using GC with the Mendelian predictions (model 6) produces a plot more closely aligned with the diagonal line, although it tends to underpredict for lower risks and overpredict for higher risks. The Oracle models (models 7 and 8) perform well in moderate calibration \citep{van2016calibration}, with model 8 which incorporates GC achieving the best calibration plot. Platt scaling (model 9)  performs well in mean and weak calibration; however, it significantly underpredicts for lower risks and significantly overpredicts for higher risks.

Although the misspecified penetrances lead to poorly calibrated Mendelian models, the model discrimination is fairly robust to this type of misspecification, as seen in the high AUC values. GB performs well in mean and weak calibration but is comparatively less strong in discrimination. The combination of GB and Mendelian models captures the best of both approaches, recalibrating the Mendelian model with misspecified penetrances while still utilizing the discriminatory power of the Mendelian predictions.

We next compare our models to the ``oracle" Mendelian models which use the true data-generating penetrances for all cancers. The oracle Mendelian models (one which uses all three cancers (model 8) and one which only uses CRC and EC (model 7) both perform well in mean and weak calibration, while the model that uses GC information naturally has better discrimination. These oracle models do not actually provide improvements in AUC and rBS over their corresponding models with misspecified CRC and EC penetrances (models 1 and 4). Instead, the main improvements are seen in the mean and weak calibration. Accordingly, our proposed GB approach, initializing with the Mendelian predictions, performs similarly to the oracle Mendelian models in the three main performance measures.

The Mendelian models in models 1, 3, 4, and 6 are all run with misspecified (different from the data-generating) CRC and EC penetrances. These misspecified penetrances represent the population in which the Mendelian models were trained, while the data-generating penetrances represent the population in which they are being applied. The Oracle Mendelian models 7 and 8 illustrate the situation where the two populations are the same. We see that model 1 has an average O/E of 0.855. Model 3, which revises model 1 (without using GC information), is able to recalibrate the predictions for the new population, resulting in an average O/E of 1.003. Thus we see evidence that GB is able to recalibrate models on populations that are different than the one used to train the original model.

We can also compare the results from the two data-generating scenarios. Overall, the Mendelian model performs worse in mean and weak calibration for the high-penetrance scenario, as the difference between the correctly and misspecified penetrances is larger in this situation. However, the increase in penetrance allows the Mendelian model to retain better discrimination and overall performance. Similarly, GB performs better in discrimination and accuracy in the high-penetrance scenario, without much change in mean calibration (as GB is unaffected by the misspecified penetrances). Thus, as expected, high-penetrance cancers provide more information for predicting the mutation carrier status, though using misspecified penetrances can affect the mean and weak calibration. Overall, the comparisons between the Mendelian and GB approaches are similar for both scenarios.

\subsubsection{Supplementary Analyses} \label{simres_supp}

We conducted additional analyses, with results provided in the supplementary materials. A discussion of these analyses are provided in Supplementary Section~\ref{ssec.supp_sim}. We explored the impact of the following parameters: the number of iterations, the extent of the GC penetrance misspecification, and the sample size. We also used a bootstrap validation approach to compare with Monte Carlo cross-validation. Lastly, we explored two situations: one where the training and testing sets are different, and another that more closely mimics the data application in Section~\ref{data}.

\begin{table}
\caption{Performance measures for the simulated data with high-penetrance data-generating CRC and EC. For each metric, we provide the mean across the 100 simulated data sets (each data set with 10,000 families), the 95\% confidence interval (CI), and the improvement percentage (IP), or the percentage of simulated data sets where the model outperformed model 1. The Mendelian model is MMRpro. The oracle Mendelian models use the data-generating CRC and EC penetrances. All gradient boosting models are run with 50 iterations.}
\label{tab_sim}

\begin{center}
\begin{scriptsize}
\begin{tabular}{lccccccccc}
  \toprule
\multirow{2}{*}[-0.5em]{Model} & \multicolumn{3}{c}{O/E} & \multicolumn{3}{c}{AUC} & \multicolumn{3}{c}{rBS} \\ 
\cmidrule(lr){2-4}\cmidrule(lr){5-7}\cmidrule(lr){8-10}
& Mean & CI & IP & Mean & CI & IP & Mean & CI & IP \\
  \midrule
  \multicolumn{10}{l}{\textbf{Low-penetrance data-generating CRC and EC}} \\
  \multicolumn{10}{l}{\textit{Without GC, using misspecified CRC and EC penetrances}} \\
  \ta & 0.855 & (0.792, 0.903) & - & 0.785 & (0.761, 0.805) & - & 0.218 & (0.211, 0.224) & - \\ 
  \tb & 0.961 & (0.948, 0.978) & 100 & 0.732 & (0.708, 0.754) & 0 & 0.228 & (0.220, 0.234) & 0 \\ 
  \tc & 1.003 & (0.990, 1.023) & 100 & 0.782 & (0.759, 0.803) & 19 & 0.218 & (0.211, 0.224) & 2 \\ 
     \addlinespace
    \multicolumn{10}{l}{\textit{With GC, using misspecified CRC and EC penetrances}} \\
    \td & 0.872 & (0.810, 0.918) & 97 & 0.847 & (0.830, 0.864) & 100 & 0.211 & (0.205, 0.217) & 100 \\ 
  \te & 0.962 & (0.949, 0.977) & 100 & 0.797 & (0.778, 0.812) & 85 & 0.224 & (0.217, 0.230) & 0 \\ 
  \tf & 1.003 & (0.990, 1.022) & 100 & 0.836 & (0.818, 0.850) & 100 & 0.215 & (0.208, 0.222) & 100 \\ 
  \addlinespace
    \multicolumn{10}{l}{\textit{Oracle Mendelian models, using correctly-specified CRC and EC penetrances}} \\
    \tg & 1.001 & (0.927, 1.058) & 99 & 0.787 & (0.763, 0.806) & 77 & 0.217 & (0.210, 0.223) & 99 \\ 
  \thh & 1.001 & (0.931, 1.054) & 99 & 0.850 & (0.832, 0.867) & 100 & 0.211 & (0.204, 0.217) & 100 \\ 
  \addlinespace
  \multicolumn{10}{l}{\textit{Other updating methods on model 1}} \\
  \ti & 1.003 & (0.990, 1.022) & 100 & 0.785 & (0.761, 0.805) & 0 & 0.220 & (0.212, 0.226) & 0 \\ 
  \tj & 1.113 & (1.094, 1.141) & 89 & 0.785 & (0.761, 0.805) & 7 & 0.218 & (0.211, 0.224) & 15 \\ 
   \midrule
  \multicolumn{10}{l}{\textbf{High-penetrance data-generating CRC and EC}} \\
    \multicolumn{10}{l}{\textit{Without GC, using misspecified CRC and EC penetrances}} \\
 \ta & 0.742 & (0.702, 0.794) & - & 0.916 & (0.907, 0.927) & - & 0.196 & (0.190, 0.203) & - \\ 
  \tb & 0.958 & (0.946, 0.973) & 100 & 0.869 & (0.855, 0.880) & 0 & 0.215 & (0.208, 0.222) & 0 \\ 
  \tc & 1.001 & (0.990, 1.015) & 100 & 0.914 & (0.904, 0.923) & 3 & 0.197 & (0.190, 0.203) & 28 \\ 
   \addlinespace
  \multicolumn{10}{l}{\textit{With GC, using misspecified CRC and EC penetrances}} \\
  \td & 0.771 & (0.730, 0.820) & 100 & 0.930 & (0.919, 0.938) & 100 & 0.193 & (0.186, 0.199) & 100 \\ 
  \te & 0.959 & (0.947, 0.973) & 100 & 0.882 & (0.869, 0.892) & 0 & 0.214 & (0.207, 0.221) & 0 \\ 
  \tf & 1.002 & (0.991, 1.015) & 100 & 0.922 & (0.913, 0.931) & 100 & 0.196 & (0.189, 0.202) & 73 \\ 
  \addlinespace
  \multicolumn{10}{l}{\textit{Oracle Mendelian models, using correctly-specified CRC and EC penetrances}} \\
   \tg & 1.004 & (0.948, 1.073) & 100 & 0.917 & (0.908, 0.928) & 96 & 0.194 & (0.187, 0.201) & 100 \\ 
  \thh & 1.002 & (0.943, 1.065) & 100 & 0.931 & (0.921, 0.939) & 100 & 0.191 & (0.184, 0.198) & 100 \\ 
  \addlinespace
  \multicolumn{10}{l}{\textit{Other updating methods on model 1}} \\
  \ti & 1.002 & (0.990, 1.014) & 100 & 0.916 & (0.907, 0.927) & 0 & 0.197 & (0.189, 0.204) & 11 \\ 
  \tj & 1.102 & (1.085, 1.123) & 100 & 0.916 & (0.907, 0.927) & 0 & 0.196 & (0.188, 0.202) & 99 \\
  \bottomrule
\end{tabular}
\end{scriptsize}
\end{center}

\begin{flushleft}
\begin{footnotesize}
$^\text{a}$ Mendelian model with misspecified CRC and EC penetrances, without GC \\
$^\text{b}$ GB without Mendelian predictions, using CRC and EC \\
$^\text{c}$ GB initialized with Mendelian predictions from model 1, using CRC and EC \\
$^\text{d}$ Mendelian model with misspecified CRC and EC penetrances, and incorporating GC through the data-generating penetrance \\
$^\text{e}$ GB without Mendelian predictions, using all 3 cancers \\
$^\text{f}$ GB initialized with Mendelian predictions from model 1, using all 3 cancers \\
$^\text{g}$ Oracle Mendelian model with data-generating CRC and EC penetrances, without GC \\
$^\text{h}$ Oracle Mendelian model with data-generating CRC and EC penetrances, and incorporating GC through the data-generating penetrance \\
\end{footnotesize}
\end{flushleft}

\end{table}

\section{Data Application} \label{data}

\subsection{USC-Stanford Data} \label{data-usc}

We apply our methods to data from the University of Southern California (USC) Norris Comprehensive Cancer Center and the Stanford Cancer Institute Cancer Genetics Hereditary Cancer Panel (HCP) study \citep{idos2018promoting}. Investigators collected these data to study the impact of several genetic mutations on cancer outcomes as well as the role of genetic testing in clinical settings. Participants were meant to be representative of a high-risk population; thus, individuals with greater than 2.5\% risk of carrying a cancer susceptibility gene mutation, as measured by various risk prediction models (including MMRpro), were considered eligible for the study. Enrollment lasted between 2014 and 2016, and in total 2000 participants were enrolled. Each participant received the Myriad Genetics myRisk Hereditary Cancer test, which identifies mutations for 25 genes shown to increase cancer susceptibility, including the three MMR genes included in MMRpro (some received Myriad's 28-gene panel, which includes an additional three genes). The panel also includes other genes which have been shown to be associated with CRC, EC, and GC, such as \textit{EPCAM} and \textit{PMS2}. Each participant completed baseline questionnaires providing information on risk factors and family history.

Of the 2000 participants, 102 had variants of uncertain significance (VUS) for at least one of the MMR genes. By definition, it is unknown whether or not VUS carriers have pathogenic or benign variants; thus, we  exclude individuals with VUS's from our analysis. Also, 204 tested negative for the three MMR mutations but positive for other mutations. As they carry mutations that lead to increased cancer susceptibility, we preferred not to consider these as negative, but they could not be considered as positive as the models are calibrated to predict carrier status for the three MMR genes only. Thus we excluded them as well. Lastly, we excluded 5 participants whose families resulted in errors when running MMRpro, due to entry errors or inbred relationships not handled by MMRpro. After these exclusions, we are left with a final subset of 1689 participants, 27 of whom tested positive for an MMR mutation. A summary of this subset is provided in Table~\ref{tab_usc}. We see that the probands in this cohort have significantly more cancer history compared to the probands in the simulated data (Supplementary Table~\ref*{stab:simfams}).

\begin{table}
\centering
\caption{Summary of the USC-Stanford data.}
\label{tab_usc}
\begin{tabular}{ll}
  \toprule
  {\bf General} & \\
  Number of families & 1689 \\
  Average family size & 33.67 \\
  Number of male participants & 325 \\
  \addlinespace
  {\bf Participant Cancer History} & \\
  \textit{Number of participants with:} & \\
  Colorectal cancer (average age) & 262 (50.15) \\
  Endometrial cancer (average age) & 65 (50.62) \\
  Gastric cancer (average age) & 40 (50.03) \\
  \addlinespace
  {\bf Family Member Cancer History} & \\
  \textit{Number of families with at least one non-participant with:} & \\
  Colorectal cancer & 463 \\
  Endometrial cancer & 227 \\
  Gastric cancer & 270 \\
  \addlinespace
  {\bf Participant MMR Mutations} & \\
  \textit{Number of participants with mutations of:} & \\
  \textit{MLH1} & 9 \\
  \textit{MSH2} & 11 \\
  \textit{MSH6} & 7 \\
  \addlinespace
  {\bf Participant Race} & \\
  \textit{Number of participants whose race is:} & \\
  Asian & 184 \\
  Black & 68 \\
  Hispanic & 661 \\
  Native American & 6 \\
  White & 691 \\
  Unknown & 79 \\
   \bottomrule
\end{tabular}
\end{table}

We ran models 1-6 and 9-10 from Section~\ref{simsetup}, with the participant as the counselee. Naturally, unlike in simulations, we do not know the true ``data-generating" penetrances, and hence we do not know if the penetrances in the Mendelian models are correctly specified. For CRC and EC, we use the MMRpro penetrance estimates, which are used in the high-penetrance scenario in the simulations. For GC, we use penetrance estimates from the literature \citep{dowty2013cancer,moller2018cancer}, with lifetime risks for non-carriers, carriers of \textit{MLH1} mutations, carriers of \textit{MSH2} mutations, and carriers of \textit{MSH6} mutations of 0.006, 0.163, 0.006, and 0.035, respectively, for males; and 0.006, 0.163, 0.201, and 0.035, respectively, for females. In addition, we use the MMRpro allele frequencies of 0.0004, 0.0005, and 0.0002 for \textit{MLH1}, \textit{MSH2}, and \textit{MSH6}, respectively. MMRpro can account for information on race, microsatellite instability (MSI) testing, and immunohistochemistry (IHC) testing. These are not explicitly referred to in our model description in Section~\ref{mendel}, but, since the data included such information, we used it to obtain our MMRpro predictions. We used the same GB tuning parameters and decision trees as in the simulations, specified in Section~\ref{simsetup}. Again we split the data into training and testing sets of equal size, doing this repeatedly with random splits to account for uncertainty in the splitting process.
As in the simulations, we provide a sensitivity analysis for the number of iterations in the GB algorithm as well as the extent of alteration in the GC penetrances, in Supplementary Tables~\ref*{stab:usc_sens} and \ref*{stab:usc_mis}.

\subsection{Results} \label{results}

A summary of the results is provided in Table~\ref{tab_usc_res}.
We first compare the three models that do not use GC information. The Mendelian model (model 1) had better discrimination and mean calibration than the GB model without Mendelian predictions (model 2). The combination of GB and the Mendelian model (model 3) had the best mean calibration on average, although it had worse O/E compared to model 1 in 71\% of the Monte Carlo cross-validation replicates. It also had worse discrimination compared to model 1.
Overall, the performance metrics varied markedly due to the small sample size and few number of carriers.

We also explored the various approaches to incorporating GC information. The completely Mendelian approach of using the GC penetrance (model 4) actually provided worse mean calibration and discrimination compared to the Mendelian model without GC information (model 1). This suggests that the estimated GC penetrance used in the Mendelian model is misspecified and not representative of the true data-generating mechanism in the data. Thus, we see a clear example of a situation where published penetrance estimates may be inaccurate and harmful to Mendelian prediction. The GB approach to incorporating information from the three cancers, without the Mendelian prediction (model 5), had similar performance measures as the equivalent GB approach without using GC as a feature (model 2). Thus we see that GC did not add to CRC and EC for predicting MMR carrier status. The combination of GB with the Mendelian prediction, using all three cancers as features (model 6), provided the best mean calibration but had worse discrimination than the Mendelian model with GC information (model 4), with similar rBS values. In addition, it performed worse in O/E and AUC compared to model 1 in a majority of the cross-validation replicates.

Although model 6 didn't seem to be a clearly better approach than the Mendelian model in model 4, we can see its effectiveness when separating families with and without a family history of GC, as shown in Table~\ref{tab_gc}. Among families with GC, Model 6 outperforms model 4 significantly in all 5 metrics. Thus, although both models incorporate GC information, the GB method led to better predictions. This illustrates the difficulty in incorporating new cancers in Mendelian models, as the GC penetrance is likely misspecified for this data set and hence led to miscalibrated predictions. In these situations, GB can be a more effective approach to incorporating this information.

\begin{figure}
\centering
\label{plot_gb_mmr}
\includegraphics[scale = 0.35]{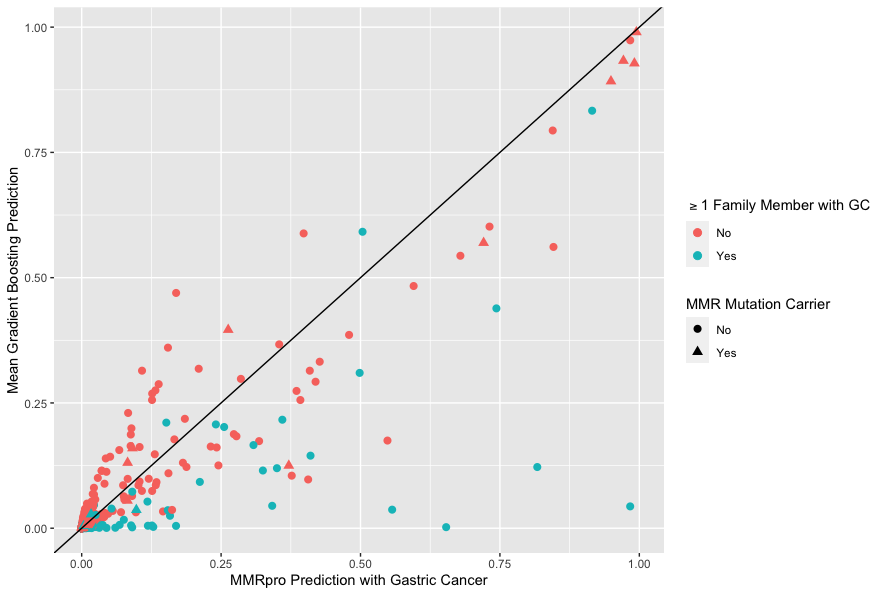}
\caption{The mean (over the Monte Carlo cross-validation replicates) gradient boosting predictions initialized by the Mendelian predictions (model 6) compared to the completely Mendelian predictions with gastric cancer (model 4) for the USC-Stanford HCP study data. The gradient boosting predictions are initialized with the MMRpro predictions without gastric cancer information (model 1) and obtained using 50 iterations. There are a total of 26 carriers (triangles) and 1670 non-carriers (circles). There are also 307 families with at least one member with gastric cancer (blue), while 1389 families do not have members with gastric cancer (red).}
\end{figure}

We can visualize the differences between these two models among families with and without a history of GC in Figure~\ref{plot_gb_mmr}. Here we plot predictions from model 4 against the average predictions from model 6 over the 1000 Monte Carlo cross-validation replicates. For each participant, we only average over the replicates where the participant was selected to be in the testing set. The plot shows the improvement in calibration for GB, as the risk predictions are lower on average compared to the completely Mendelian model with GC. In particular, we notice that several non-carriers had high predicted risks using the completely Mendelian model with GC, while the GB predictions were much lower. Among these non-carriers, many of the ones with the largest differences in risk predictions were also in families with at least one member with GC, showing how GB may handle GC information better than the Mendelian model with the current GC penetrance estimate.

We can assess weak calibration using the calibration slopes and intercepts in Supplementary Table~\ref*{stab:cal_usc}. The GB with Mendelian models (models 3 and 6) on average had the best mean calibration among models 1-6, as measured using the intercept. However, they had intercepts further from 0 compared to model 1 for a majority of the cross-validation replicates. When measuring weak calibration using the calibration slopes, both models 3 and 6 outperformed model 1 for a large majority of the cross-validation replicates. Platt scaling (model 9) outperformed the GB with Mendelian models in both mean and weak calibration. The calibration plots are also provided in Supplementary Figure~\ref*{sfig:cal_usc}. We see that all the models struggle in moderate calibration, as the plots deviate significantly from the diagonal line.

The impact of the number of iterations is shown in Supplementary Table~\ref*{stab:usc_sens} and was similar to that seen in simulations. The average O/E values for GB initialized with the Mendelian predictions was fairly stable, although the O/E ratios did increase as the number of iterations increased, and the confidence intervals were wide. The AUC values also decreased as the number of iterations increased, while the rBS values remained similar. The O/E ratios for GB without the Mendelian predictions increased dramatically as the number of iterations increased, while the AUC and rBS values changed slightly.

As in the simulations, the Mendelian model calibration was sensitive to alterations to the GC penetrance, as seen in Supplementary Table~\ref*{stab:usc_mis}. The O/E increased with the exponent used to modify the GC survival function. All levels of alterations resulted in overprediction on average, as the O/E ratio was 0.907 when using the GC survival function with an exponent of 4. This suggests that the GC penetrance from ASK2ME corresponds to substantially higher GC risks than the penetrance underlying this study.

Overall, although the mean performance measures seem to provide evidence of these assertions, all the confidence intervals were much wider than the ones from the simulations, constraining our ability to arrive at definitive conclusions. Compared to the simulated data, we have a smaller data set with fewer mutation carriers. 

\begin{table}
\caption{Performance measures for the USC-Stanford HCP study data, with and without gastric cancer (GC) information. For each metric, we provide the mean across the 1000 Monte Carlo cross-validation replicates, the 95\% confidence interval (CI), and the improvement percentage (IP), or the percentage of Monte Carlo cross-validation replicates where the model outperformed model 1. The Mendelian model used is MMRpro. All gradient boosting models are run with 50 iterations.}
\label{tab_usc_res}

\begin{center}
\begin{footnotesize}
\begin{tabular}{lccccccccc}
  \toprule
\multirow{2}{*}[-0.5em]{Model} & \multicolumn{3}{c}{O/E} & \multicolumn{3}{c}{AUC} & \multicolumn{3}{c}{rBS} \\ 
\cmidrule(lr){2-4}\cmidrule(lr){5-7}\cmidrule(lr){8-10}
& Mean & CI & IP & Mean & CI & IP & Mean & CI & IP \\
  \midrule
  \multicolumn{10}{l}{\textit{Without GC}} \\
  \ta & 0.869 & (0.546, 1.236) & 0 & 0.862 & (0.817, 0.906) & 0 & 0.132 & (0.110, 0.152) & 0 \\ 
  \tb & 0.809 & (0.384, 1.441) & 22 & 0.718 & (0.626, 0.810) & 0 & 0.125 & (0.103, 0.144) & 86 \\ 
  \tc & 0.991 & (0.424, 2.034) & 29 & 0.814 & (0.682, 0.893) & 12 & 0.129 & (0.111, 0.147) & 76 \\ 
    \addlinespace
    \multicolumn{10}{l}{\textit{With GC}} \\
  \td & 0.759 & (0.465, 1.096) & 12 & 0.844 & (0.793, 0.896) & 0 & 0.138 & (0.118, 0.158) & 1 \\ 
 \te & 0.807 & (0.377, 1.441) & 21 & 0.716 & (0.620, 0.809) & 0 & 0.125 & (0.104, 0.144) & 86 \\ 
  \tf & 0.985 & (0.419, 1.941) & 27 & 0.815 & (0.685, 0.895) & 14 & 0.129 & (0.111, 0.148) & 75 \\ 
  \addlinespace
  \multicolumn{10}{l}{\textit{Other updating methods on model 1}} \\
   \ti & 1.047 & (0.460, 2.047) & 37 & 0.862 & (0.817, 0.906) & 0 & 0.120 & (0.097, 0.142) & 97 \\ 
  \tj & 1.579 & (0.515, 5.030) & 32 & 0.862 & (0.817, 0.906) & 4 & 0.120 & (0.096, 0.143) & 98 \\ 
  \bottomrule
\end{tabular}
\end{footnotesize}
\end{center}
\vspace{1mm}

\begin{flushleft}
$^\text{a}$ Mendelian model, without gastric cancer information \\
$^\text{b}$ Gradient boosting initialized without Mendelian predictions, using colorectal and endometrial cancer information \\
$^\text{c}$ Gradient boosting initialized with Mendelian predictions from (1), using colorectal and endometrial cancer information \\
$^\text{d}$ Mendelian model, incorporating gastric cancer information through the penetrance \\
$^\text{e}$ Gradient boosting without Mendelian predictions, using all 3 cancers \\
$^\text{f}$ Gradient boosting initialized with Mendelian predictions from model 1, using all 3 cancers \\
\end{flushleft}

\end{table}

\begin{table}
\caption{Comparison of the performance metrics for gradient boosting initialized by the Mendelian predictions from model 1, using all three cancers as features (model 6) and a completely Mendelian model with gastric cancer information (model 4) in the HCP data. We compare the two models, subsetted by families who have or do not have at least one member with gastric cancer. For each metric, we provide the mean across the 1000 Monte Carlo cross-validation replicates, the 95\% confidence interval (CI), and the improvement percentage (IP), or the percentage of Monte Carlo cross-validation replicates where the model outperformed model 1.}
\label{tab_gc}

\begin{center}
\begin{footnotesize}
\begin{tabular}{rlll}
  \toprule
  & & (4) Mend. with GC & (6) GB with Mend. \\ \hline
  \multicolumn{3}{l}{\textit{GC in family}} \\
\multirow{3}{*}{O/E} & Mean & 0.278 & 0.846 \\ 
  & CI & (0.000, 0.673) & (0.000, 3.257) \\ 
  & IP & - & 75 \\ \addlinespace
  \multirow{3}{*}{AUC} & Mean & 0.823 & 0.897 \\ 
  & CI & (0.696, 0.933) & (0.764, 0.962) \\ 
  & IP & - & 97 \\ \addlinespace
  \multirow{3}{*}{rBS} & Mean & 0.164 & 0.121 \\ 
  & CI & (0.110, 0.206) & (0.046, 0.169) \\ 
  & IP & - & 99 \\ \addlinespace
  \multirow{3}{*}{Intercept} & Mean & -5.182 & -3.416 \\ 
  & CI & (-25.487, -0.619) & (-23.577, 1.534) \\ 
  & IP & - & 93 \\ \addlinespace
  \multirow{3}{*}{Slope} & 0.239 & 0.382 \\ 
  & CI & (0.000, 0.502) & (0.000, 0.757) \\ 
  & IP & - & 91 \\ \addlinespace
  \midrule
  \multicolumn{3}{l}{\textit{No GC in family}} \\
  \multirow{3}{*}{O/E} & Mean & 0.988 & 1.041 \\ 
  & CI & (0.598, 1.424) & (0.420, 2.067) \\ 
  & IP & - & 14 \\ \addlinespace
  \multirow{3}{*}{AUC} & Mean & 0.856 & 0.802 \\ 
  & CI & (0.803, 0.907) & (0.660, 0.891) \\ 
  & IP & - & 13 \\ \addlinespace
  \multirow{3}{*}{rBS} & Mean & 0.131 & 0.130 \\ 
  & CI & (0.106, 0.154) & (0.108, 0.151) \\ 
  & IP & - & 61 \\ \addlinespace
  \multirow{3}{*}{Intercept} & Mean & -0.071 & -0.072 \\ 
  & CI & (-0.965, 0.613) & (-1.450, 1.184) \\ 
  & IP & - & 17 \\ \addlinespace
  \multirow{3}{*}{Slope} & Mean & 0.414 & 0.452 \\ 
  & CI & (0.303, 0.525) & (0.305, 0.624) \\ 
  & IP & - & 87 \\ 
   \bottomrule
\end{tabular}
\end{footnotesize}
\end{center}
\end{table}

\section{Discussion} \label{discussion}

We explored the concept of developing a machine learning technique for improving a reference model, to address the increasingly common situation where improving is more efficient or expedient than retraining.
The general concept we championed is one where the reference model is highly complex and is modified by operating on residuals with a simpler modeling approach--in our case GB. Our approach allows us to empirically incorporate features, both ones already used in the existing model as well as new ones, while taking advantage of the strengths of the existing model. We illustrated the model improvement approach by applying it to Mendelian models, which are complex and require accurate cancer penetrance estimates as an input. Obtaining accurate penetrances is not always feasible, and results from the simulations show that when the penetrances are misspecified, Mendelian models do not provide well-calibrated predictions. In addition, while MMRpro is open-source, provided through the BayesMendel R package, other Mendelian models such as BOADICEA are proprietary and hence difficult to enhance by users. Results from simulated data provide evidence that our proposed model improvement approach for integrating cancer information improves the Mendelian model mean, weak, and moderate calibration compared to utilizing the same information using a completely Mendelian approach as well as compared to using GB without the benefit of the reference Mendelian model predictions. In addition, results from data from the USC-Stanford Genetics HCP study showed improvements in mean calibration.

Since the main improvements in our proposed gradient boosting model were found in the calibration, we compared our approach to Platt scaling and isotonic regression, two existing recalibration methods. Although these methods improved mean calibration, we saw that they did not perform strongly in weak and moderate calibration as well as discrimination compared to our approach.

We found that GB performed well in the scenario with low-penetrance data-generating CRC and EC but comparatively worse when these cancers were generated with high penetrances. This shows that extending the model via GB is especially valuable when the existing model's features are less informative for predicting the outcome. This is the case in the low penetrance scenario, where we have less cancer history that can predict the carrier status. In these situations, GB can incorporate features unused in the existing model that may be more informative, such as GC in our application.

From the simulation results, we found that GB is able to recalibrate the existing model on a new population, different than the population on which the existing model was trained, while incorporating new features to improve discrimination. Transportability to a third population, on which neither GB nor the existing model has been trained, remains a challenge. This holds not just for GB, but for model extensions in general, and illustrates the difficulty in applying prediction models on populations that are different than the populations on which the models were trained \citep{bernau2014cross}.

The most prominent improvement illustrated in our real data analysis is the correction of mean and weak miscalibration, particularly among families with a history of GC. In biomedical applications, miscalibration is common when applying a model developed on one population to another. Our model improvement mechanism is relatively simple in this case, as the miscalibration is reflected in the predictions, and GB can identify both the overall shifts in predicted values and the feature-dependent adjustments that can remove them. The improvement among families with a history of GC also demonstrates a situation where incorporating new features in the method of the original model (via penetrances in the case of Mendelian models) is difficult due to limitations in scientific knowledge. GB therefore provides an alternative, empirical approach to incorporating the information.

Although the data application results are promising, there are limitations in the data that suggest caution in the interpretation. Since there are only 25 MMR mutation carriers, the confidence intervals for the performance measures in Table~\ref{tab_usc_res} are wide, and the improvement percentages over model 1 are low. In particular, we see that even though the O/E ratios for GB initialized with the Mendelian predictions are much closer to 1 on average compared to the Mendelian model (model 1), these ratios were further from 1 compared to model 1 for a majority of the Monte Carlo cross-validation replicates. The low number of carriers is especially concerning since the algorithm was trained by splitting the data into training and testing sets, so validation on the testing set only considers 13 carriers on average. Some splits may have a disproportionate amount of carriers in either the training or testing sets, leading to unstable risk predictions. The comparatively higher effectiveness of GB-based model improvement in the simulated data is due to several factors. In the simulations we generated carrier status using allele frequencies of 0.01, which provides a significant amount of carriers. In addition, we scaled the GC penetrances to artificially create a wider gap in GC risk between mutation carriers and non-carriers; consequently, we improved the predictive power of the presence of GC in the family, which helped GB improve the predictions. We also used misspecified CRC and EC penetrances when running the Mendelian models, which reduced their predictive performance. Lastly, we used a larger sample size of 10,000 families per simulated data set and generated 100 different data sets, allowing us to obtain tight confidence intervals.

All models applied to the USC-Stanford Genetics HCP study data tended to overestimate carrier probabilities. This may be because the participants are selected to be likely to have a genetic predisposition to cancer. If there is heterogeneity in penetrance, these families could reflect higher penetrances than those used in MMRpro, which are derived from a meta-analysis only including studies that adjust for ascertainment \citep{chen2006prediction}, and hence likely better represent population-level risk. This could explain some of the overall miscalibration. In addition,the Myriad Genetics panel only considers 25 genes, ignoring a large set of less commonly mutated additional cancer susceptibility genes. It is possible that some of the individuals in the data who tested negative for all 25 genes were carriers of a cancer susceptibility gene that was not included in the panel. Since the families were selected based on their cancer history, there may be a genetic component explaining the cancer history that is not captured by the gene panel testing results. While we could exclude carriers of a non-MMR mutation included in the panel, we could not do the same for genes not included in the panel. Validating our model improvement approach on data with (1) genetic testing results for more genes and (2) a larger sample size would help better determine our model's effectiveness in this specific context.

We used a relatively basic GB model, as we only included three features in each base learner, one for each of the three cancers of interest. This had the advantage of allowing for direct comparison of the improved and reference models when they used the same features. However, including additional features could aid GB's ability to improve prediction of the carrier status. For example, we could include information from other cancers or separate the cancer information based on relationship type. A family member with cancer who is not a blood relative of the study participant does not carry as much significance as a first-degree relative with cancer. This distinction is incorporated in Mendelian models, but GB may be able to detect residual non-Mendelian signal of the cancers. We could also incorporate other risk factors besides cancer history. The USC-Stanford data includes information about aspirin use, which has been shown to be associated with a decreased risk of CRC \citep{flossmann2007effect}. Including this information in the base learner could improve the model's performance, in general and particularly if the effect of aspirin differs between carriers and others.

Overall, improving model performance while keeping the revisions simple is an important issue. Choices such as new features to incorporate, tuning parameters, loss functions, etc. are often specific to the context of the prediction model. In general, when choosing new features, we suggest using features that are both highly associated with the outcome and easily obtainable in real data.
We fixed the GB tuning parameters (the maximum tree depth, bagging fraction, and shrinkage parameter) based on recommended values from the literature. It is possible that other values may impact the results. In addition, other binary loss functions could be used to train the GB algorithm.

Our model improvement strategy uses GB, an ensemble method, where multiple prediction models are combined to increase predictive power. Other examples of ensemble methods include bagging \citep{breiman1996bagging} and stacking \citep{wolpert1992stacked}. Our approach has some points of contact with ensembling, but it may be worth pointing out that
the final predictions are not ensemble predictions from the reference model and some other model. They are an ensemble of GB weak learners trained on residuals from the reference model. The reference model does not enter linearly in the final predictions.

In sum, extending existing classification and prediction models via machine learning techniques designed to improve existing models rather than training them from scratch is an important and underutilized approach. We explored a proof-of-principle implementation and used it on Mendelian models. We see that GB can be applied as a convenient and powerful tool to extend models, leveraging the advantages of the reference model while providing opportunities for improvements through new data and features. We hope that this work will encourage others to further apply and develop methods using this framework to upgrade complex existing prediction models.

\section*{Acknowledgements}

This work was supported by the NCI at the NIH Grants 5T32CA009337-32, 4P30CA006516-51 and T32CA009001; Myriad Genetics, Inc.; NIH Grant Awards KL2TR000131 and P30CA014089; the Anton B. Burg Foundation; the Jane \& Kris Popovich Chair in Cancer Research; and a gift from Daniel and Maryann Fong.

\bibliographystyle{biom}

\bibliography{absref}

\newpage

\beginsupplement

\begin{center}
    \Huge{Supplementary Materials}
\end{center}

\noindent R code is provided at \url{https://github.com/theohuang/Gradient-boosting}.

\section{Misspecified Penetrances}
\label{ssec.mis}

In Section~\ref*{simsetup}, we describe simulation scenarios using misspecified penetrances in the Mendelian model. Here we explain the derivation of the misspecified penetrances.

Let $T_r$ be the age of the $r$th cancer type, $U$ be an indicator denoting whether the individual is male, and $\G$ be the genotype. Then for age $t$, the penetrance is defined as $P(T_r = t | \G, U)$ (we condition on $U$ because the penetrances in the Mendelian model, MMRpro, are sex-specific). The survival function is defined as $P(T_r > t | \G, U) = 1 - \sum_{s = 1}^t P(T_r = s | \G, U)$. If $P(T_r = t | \G, U)$ and $P(T_r > t | \G, U)$ are the data-generating penetrance and survival, respectively, then the misspecified survival used for the Mendelian model is $\sqrt{P(T_r > t | \G, U)}$ and the misspecified penetrance is $\sqrt{P(T_r > t - 1 | \G, U)} - \sqrt{P(T_r > t | \G, U)}$.

\section{Supplementary Simulation Analyses}
\label{ssec.supp_sim}

In Supplementary Table~\ref{stab:sim_sens}, we also explore the impact of the number of iterations of the GB on performance. This parameter controls the importance of the initial model and could play a different role in our method compared to standard applications of GB. Besides 50 used in the main analyses, we ran the GB models with 25 and 100 iterations. These additional models are displayed in Supplementary Table~\ref{stab:mod}. Model 6 with 25 and 100 iterations is called model 11 and 12, respectively, and model 5 with 25 and 100 iterations is called model 13 and 14, respectively. For the GB and Mendelian model combination, our three choices of the number of iterations led to similar performance measures, demonstrating a general robustness to the number of iterations. However, the number of iterations affected the O/E ratios for GB without the Mendelian predictions, with the O/E increasing as the number of iterations increased. Without a base prediction with which to initialize, GB struggled to provide well-calibrated results without enough iterations. The AUC and rBS values did not change based on the number of iterations. For all three choices of the number of iterations, initializing with the Mendelian predictions provided better discrimination, emphasizing the importance of the Mendelian component.

\begin{figure}
\centering
\includegraphics[scale=0.35]{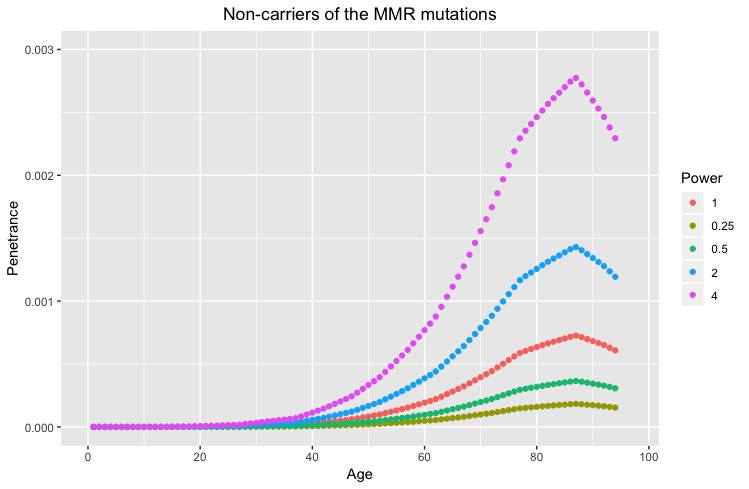}
\includegraphics[scale=0.35]{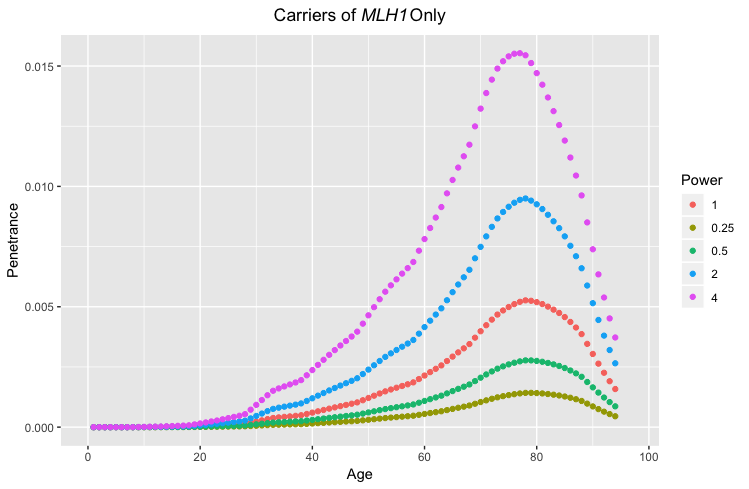}
\label{sfig_pen_gc}
\caption{Plots of the altered male gastric cancer penetrances where the survival functions are raised to different exponents. Note that when the exponent is 1, the penetrance is unchanged and hence correctly-specified.}
\end{figure}

When incorporating GC through the completely Mendelian approach, we used the data-generating GC penetrance. However, in practice the GC penetrance used in the Mendelian model may be misspecified. We explore the impact of this misspecification in Supplementary Table~\ref{stab:sim_mis}. The results show that the O/E ratios tend to decrease as the estimated penetrance is decreased and increase as the estimated penetrance is increased (see also Supplementary Figure~\ref{sfig_pen_gc}), while the AUC and rBS levels remain similar. Externally determined penetrances have been a traditional strength of Mendelian models, as they allow incorporation of external evidence. When this evidence is less reliable, however, this approach comes with the danger of impairing the calibration, compared to the Mendelian model without GC information (model 1). GB, on the other hand, does not rely directly on these penetrance estimates and can provide well-calibrated predictions.

The role of the sample size is assessed in Supplementary Table~\ref{stab:sss}. Overall, the mean performance metrics are similar when using sample sizes of 1,000 families and 10,000 families. However, the confidence intervals of the performance metrics for the smaller sample size are wider, due to the smaller number of carriers. In addition, the proportions of simulated data sets in which model outperforms model 1 are less extreme, with fewer proportions of 0 and 1. Supplementary Table~\ref{stab:boot} also provide performance metrics when using bootstrap validation instead of Monte Carlo cross-validation. Overall, the results are similar to the results obtained when using Monte Carlo cross-validation.

Supplementary Table~\ref{stab:trans} provides results when the training data is generated using high-penetrance CRC and EC and the testing data is generated using low-penetrance CRC and EC. In this scenario, as expected O/E's for all 10 models were all significantly above 1. In addition, the AUC values were significantly lower than the corresponding values for the high-penetrance scenario, shown in Table~\ref*{tab_sim}. This indicates that although GB performs well in calibrating an existing model to a new population (different than the population used to train the existing model), it does not have information to fully address transportability to a third population, unless it is re-trained on a subset of that population. Despite this limitation, GB can still potentially improve model discrimination in multiple populations due to its ability to incorporate new features.

Lastly, Supplementary Table~\ref{stab:lgc} provide results when using the smaller sample size of 1,000 families and generating data using low-penetrance GC. Here we see that while GB still improves weak calibration, it performs worse in discrimination. We note in particular that incorporating GC information does not make a noticeable difference in performance, as there are few GC cases in the data. This may explain the lack of improvement in discrimination in Section~\ref*{results} in the data application.

\newpage

\section{Table Legend (for most tables in the Supplementary materials)} \label{sleg}%

\text{} \\
\noindent $^\text{a}$ Mendelian model with misspecified CRC and EC penetrances, without GC information \\
$^\text{b}$ GB without Mendelian predictions, using CRC and EC information \\
$^\text{c}$ GB initialized with Mendelian predictions from model 1, using CRC and EC information \\
$^\text{d}$ Mendelian model with misspecified CRC and EC penetrances, and incorporating GC information through the data-generating penetrance \\
$^\text{e}$ GB without Mendelian predictions, using all 3 cancers \\
$^\text{f}$ GB initialized with Mendelian predictions from model 1, using all 3 cancers \\
$^\text{g}$ Oracle Mendelian model with data-generating CRC and EC penetrances, without GC information \\
$^\text{h}$ Oracle Mendelian model with data-generating CRC and EC penetrances, and incorporating GC information through the data-generating penetrance \\

\begin{landscape}
\begin{table}
\caption{List and description of all the prediction models used in the simulations.}
\label{stab:mod}
\begin{scriptsize}
\begin{center}
\begin{tabular}{lllllllllllllllll}
\toprule
\multirow{4}{*}{Num$^*$} & \multirow{4}{*}{Model$^\dagger$} & CRC & CRC & CRC & EC & EC & EC & GC & GC & GC & DG & DG & DG & GB & Num & CV/ \\
& & Pen$^\ddagger$ & Pen & Feat$^\mathsection$ & Pen & Pen & Feat & Pen & Pen & Feat & CRC/EC & CRC/EC & GC & Iter$^\|$ & Fam$^{***}$ & Boot$^{\dagger\dagger\dagger}$ \\
& & & Exp$^\mathparagraph$ & & & Exp & & & Exp & & Pen & Pen & Pen$^{\ddagger\ddagger}$ & & \\
& & &  & & & & & & & & Train$^{**}$ & Test$^{\dagger\dagger}$ & & & \\ \midrule
(1) & Mend & Mis & 0.5 & No  & Mis & 0.5 & No  & No  &      & No  & Low & Low & High & & 10,000 & CV    \\
(2) & GB   & No  &     & Yes & No  &     & Yes & No  &      & No  & Low & Low & High  & 50 & 10,000 & CV \\
(3) & Mend/GB & Mis & 0.5 & Yes & Mis & 0.5 & Yes & No  &      & No  & Low & Low & High & 50 & 10,000 & CV \\
(4) & Mend & Mis & 0.5 & No  & Mis & 0.5 & No  & Cor & 1    & No  & Low & Low & High & & 10,000 & CV    \\
(5) & GB   & No  &     & Yes & No  &     & Yes & No  &      & Yes & Low & Low & High  & 50 & 10,000 & CV \\
(6) & Mend/GB & Mis & 0.5 & Yes & Mis & 0.5 & Yes & Cor & 1    & Yes & Low & Low & High & 50 & 10,000 & CV \\
(7) & Mend & Cor & 1   & No  & Cor & 1   & No  & No  &      & No  & Low & Low & High & & 10,000 & CV    \\
(8) & Mend & Cor & 1   & No  & Cor & 1   & No  & Cor & 1    & No  & Low & Low & High & & 10,000 & CV    \\
(9) & Mend/Platt & Mis & 0.5 & No & Mis & 0.5 & No  & No  &      & No  & Low & Low & High & & 10,000 & CV \\
(10) & Mend/IR & Mis & 0.5 & No  & Mis & 0.5 & No  & No  &      & No  & Low & Low & High & & 10,000 & CV \\
(1)    & Mend & Mis & 0.5 & No  & Mis & 0.5 & No  & No  &      & No  & High & High & High &  & 10,000 & CV   \\
(2)    & GB   & No  &     & Yes & No  &     & Yes & No  &      & No  & High & High & High & 50 & 10,000 & CV \\
(3)    & Mend/GB & Mis & 0.5 & Yes & Mis & 0.5 & Yes & No  &      & No  & High & High & High & 50 & 10,000 & CV  \\
(4)    & Mend & Mis & 0.5 & No  & Mis & 0.5 & No  & Cor & 1    & No  & High & High & High &  & 10,000 & CV   \\
(5)    & GB   & No  &     & Yes & No  &     & Yes & No  &      & Yes & High & High & High & 50 & 10,000 & CV \\
(6)    & Mend/GB & Mis & 0.5 & Yes & Mis & 0.5 & Yes & Cor & 1    & Yes & High & High & High & 50 & 10,000 & CV \\
(7)    & Mend & Cor & 1   & No  & Cor & 1   & No  & No  &      & No  & High & High & High & & 10,000 & CV    \\
(8)    & Mend & Cor & 1   & No  & Cor & 1   & No  & Cor & 1    & No  & High & High & High &  & 10,000 & CV   \\
(9) & Mend/Platt & Mis & 0.5 & No & Mis & 0.5 & No  & No  &      & No  & High & High & High & & 10,000 & CV \\
(10) & Mend/IR & Mis & 0.5 & No  & Mis & 0.5 & No  & No  &      & No  & High & High & High & & 10,000 & CV \\
\bottomrule
\end{tabular}
\end{center}
\begin{flushleft}
$^*$ Model number referenced in main text \\
$^\dagger$ Mendelian model (Mend), gradient boosting (GB), Mendelian/GB combination (Mend/GB), Mendelian/Platt scaling combination (Mend/Platt), Mendelian/Isotonic regression combination (Mend/IR) \\
$^\ddagger$ Does the model use a CRC penetrance? If yes, is the penetrance misspecified (Mis) or correctly specified (Cor)? The ``EC Penet" and ``GC Penet" columns are defined analogously. \\
$^\mathparagraph$ The exponent applied to the CRC survival function in the model, if the model uses a CRC penetrance. The ``EC Penet Exponent" and ``GC Penet Exponent" columns are defined analogously. \\
$^\mathsection$ Does the model use CRC information as a feature? The ``EC Feature" and ``GC Feature" columns are defined analogously. \\
$^{**}$ Data-generating CRC and EC penetrances for the training set -- can be low-penetrance (Low) or high-penetrance (High) \\
$^{\dagger\dagger}$ Data-generating CRC and EC penetrances for the testing set -- can be low-penetrance (Low) or high-penetrance (High) \\
$^\|$ Number of iterations in the gradient boosting algorithm
$^{***}$ Number of families in the data set (training and testing combined) \\
$^{\dagger\dagger\dagger}$ Monte Carlo cross-validation (CV) or bootstrap (Boot) approach
\end{flushleft}
\end{scriptsize}
\end{table}

\begin{table}
\begin{scriptsize}
\begin{center}
Table S1 (cont.)
\begin{tabular}{lllllllllllllllll}
\toprule
\multirow{4}{*}{Num$^*$} & \multirow{4}{*}{Model$^\dagger$} & CRC & CRC & CRC & EC & EC & EC & GC & GC & GC & DG & DG & DG & GB & Num & CV/ \\
& & Pen$^\ddagger$ & Pen & Feat$^\mathsection$ & Pen & Pen & Feat & Pen & Pen & Feat & CRC/EC & CRC/EC & GC & Iter$^\|$ & Fam$^{***}$ & Boot$^{\dagger\dagger\dagger}$ \\
& & & Exp$^\mathparagraph$ & & & Exp & & & Exp & & Pen & Pen & Pen$^{\ddagger\ddagger}$ & & \\
& & &  & & & & & & & & Train$^{**}$ & Test$^{\dagger\dagger}$ & & & \\ \midrule
(1) & Mend & Mis & 0.5 & No  & Mis & 0.5 & No  & No  &      & No  & Low & Low & High & & 1,000 & CV     \\
(2) & GB   & No  &     & Yes & No  &     & Yes & No  &      & No  & Low & Low & High  & 50 & 1,000 & CV \\
(3) & Mend/GB & Mis & 0.5 & Yes & Mis & 0.5 & Yes & No  &      & No  & Low & Low & High & 50 & 1,000 & CV \\
(4) & Mend & Mis & 0.5 & No  & Mis & 0.5 & No  & Cor & 1    & No  & Low & Low & High & & 1,000 & CV    \\
(5) & GB   & No  &     & Yes & No  &     & Yes & No  &      & Yes & Low & Low & High  & 50 & 1,000 & CV \\
(6) & Mend/GB & Mis & 0.5 & Yes & Mis & 0.5 & Yes & Cor & 1    & Yes & Low & Low & High & 50 & 1,000 & CV \\
(7) & Mend & Cor & 1   & No  & Cor & 1   & No  & No  &      & No  & Low & Low & High & & 1,000 & CV    \\
(8) & Mend & Cor & 1   & No  & Cor & 1   & No  & Cor & 1    & No  & Low & Low & High & & 1,000 & CV    \\
(9) & Mend/Platt & Mis & 0.5 & No & Mis & 0.5 & No  & No  &      & No  & Low & Low & High & & 1,000 & CV \\
(10) & Mend/IR & Mis & 0.5 & No  & Mis & 0.5 & No  & No  &      & No  & Low & Low & High & & 1,000 & CV \\
(1)    & Mend & Mis & 0.5 & No  & Mis & 0.5 & No  & No  &      & No  & High & High & High &  & 1,000 & CV   \\
(2)    & GB   & No  &     & Yes & No  &     & Yes & No  &      & No  & High & High & High & 50 & 1,000 & CV \\
(3)    & Mend/GB & Mis & 0.5 & Yes & Mis & 0.5 & Yes & No  &      & No  & High & High & High & 50 & 1,000 & CV  \\
(4)    & Mend & Mis & 0.5 & No  & Mis & 0.5 & No  & Cor & 1    & No  & High & High & High &  & 1,000 & CV   \\
(5)    & GB   & No  &     & Yes & No  &     & Yes & No  &      & Yes & High & High & High & 50 & 1,000 & CV \\
(6)    & Mend/GB & Mis & 0.5 & Yes & Mis & 0.5 & Yes & Cor & 1    & Yes & High & High & High & 50 & 1,000 & CV \\
(7)    & Mend & Cor & 1   & No  & Cor & 1   & No  & No  &      & No  & High & High & High & & 1,000 & CV    \\
(8)    & Mend & Cor & 1   & No  & Cor & 1   & No  & Cor & 1    & No  & High & High & High &  & 1,000 & CV   \\
(9) & Mend/Platt & Mis & 0.5 & No & Mis & 0.5 & No  & No  &      & No  & High & High & High & & 1,000 & CV \\
(10) & Mend/IR & Mis & 0.5 & No  & Mis & 0.5 & No  & No  &      & No  & High & High & High & & 1,000 & CV \\
(1) & Mend & Mis & 0.5 & No  & Mis & 0.5 & No  & No  &      & No  & Low & Low & High & & 10,000 & Boot     \\
(2) & GB   & No  &     & Yes & No  &     & Yes & No  &      & No  & Low & Low & High  & 50 & 10,000 & Boot \\
(3) & Mend/GB & Mis & 0.5 & Yes & Mis & 0.5 & Yes & No  &      & No  & Low & Low & High & 50 & 10,000 & Boot \\
(4) & Mend & Mis & 0.5 & No  & Mis & 0.5 & No  & Cor & 1    & No  & Low & Low & High & & 10,000 & Boot    \\
(5) & GB   & No  &     & Yes & No  &     & Yes & No  &      & Yes & Low & Low & High  & 50 & 10,000 & Boot \\
(6) & Mend/GB & Mis & 0.5 & Yes & Mis & 0.5 & Yes & Cor & 1    & Yes & Low & Low & High & 50 & 10,000 & Boot \\
(7) & Mend & Cor & 1   & No  & Cor & 1   & No  & No  &      & No  & Low & Low & High & & 10,000 & Boot    \\
(8) & Mend & Cor & 1   & No  & Cor & 1   & No  & Cor & 1    & No  & Low & Low & High & & 10,000 & Boot    \\
(9) & Mend/Platt & Mis & 0.5 & No & Mis & 0.5 & No  & No  &      & No  & Low & Low & High & & 10,000 & Boot \\
(10) & Mend/IR & Mis & 0.5 & No  & Mis & 0.5 & No  & No  &      & No  & Low & Low & High & & 10,000 & Boot \\
(1)    & Mend & Mis & 0.5 & No  & Mis & 0.5 & No  & No  &      & No  & High & High & High &  & 10,000 & Boot   \\
(2)    & GB   & No  &     & Yes & No  &     & Yes & No  &      & No  & High & High & High & 50 & 10,000 & Boot \\
(3)    & Mend/GB & Mis & 0.5 & Yes & Mis & 0.5 & Yes & No  &      & No  & High & High & High & 50 & 10,000 & Boot  \\
(4)    & Mend & Mis & 0.5 & No  & Mis & 0.5 & No  & Cor & 1    & No  & High & High & High &  & 10,000 & Boot   \\
(5)    & GB   & No  &     & Yes & No  &     & Yes & No  &      & Yes & High & High & High & 50 & 10,000 & Boot \\
(6)    & Mend/GB & Mis & 0.5 & Yes & Mis & 0.5 & Yes & Cor & 1    & Yes & High & High & High & 50 & 10,000 & Boot \\
(7)    & Mend & Cor & 1   & No  & Cor & 1   & No  & No  &      & No  & High & High & High & & 10,000 & Boot    \\
(8)    & Mend & Cor & 1   & No  & Cor & 1   & No  & Cor & 1    & No  & High & High & High &  & 10,000 & Boot   \\
(9) & Mend/Platt & Mis & 0.5 & No & Mis & 0.5 & No  & No  &      & No  & High & High & High & & 10,000 & Boot \\
(10) & Mend/IR & Mis & 0.5 & No  & Mis & 0.5 & No  & No  &      & No  & High & High & High & & 10,000 & Boot \\
\bottomrule
\end{tabular}
\end{center}
\end{scriptsize}
\end{table}

\begin{table}
\begin{scriptsize}
\begin{center}
Table S1 (cont.)
\begin{tabular}{lllllllllllllllll}
\toprule
\multirow{4}{*}{Num$^*$} & \multirow{4}{*}{Model$^\dagger$} & CRC & CRC & CRC & EC & EC & EC & GC & GC & GC & DG & DG & DG & GB & Num & CV/ \\
& & Pen$^\ddagger$ & Pen & Feat$^\mathsection$ & Pen & Pen & Feat & Pen & Pen & Feat & CRC/EC & CRC/EC & GC & Iter$^\|$ & Fam$^{***}$ & Boot$^{\dagger\dagger\dagger}$ \\
& & & Exp$^\mathparagraph$ & & & Exp & & & Exp & & Pen & Pen & Pen$^{\ddagger\ddagger}$ & & \\
& & &  & & & & & & & & Train$^{**}$ & Test$^{\dagger\dagger}$ & & & \\ \midrule
(1) & Mend & Mis & 0.5 & No  & Mis & 0.5 & No  & No  &      & No  & High & Low & High & & 1,000 & CV     \\
(2) & GB   & No  &     & Yes & No  &     & Yes & No  &      & No  & High & Low & High  & 50 & 1,000 & CV \\
(3) & Mend/GB & Mis & 0.5 & Yes & Mis & 0.5 & Yes & No  &      & No  & High & Low & High & 50 & 1,000 & CV \\
(4) & Mend & Mis & 0.5 & No  & Mis & 0.5 & No  & Cor & 1    & No  & High & Low & High & & 1,000 & CV    \\
(5) & GB   & No  &     & Yes & No  &     & Yes & No  &      & Yes & High & Low & High  & 50 & 1,000 & CV \\
(6) & Mend/GB & Mis & 0.5 & Yes & Mis & 0.5 & Yes & Cor & 1    & Yes & High & Low & High & 50 & 1,000 & CV \\
(7) & Mend & Cor & 1   & No  & Cor & 1   & No  & No  &      & No  & High & Low & High & & 1,000 & CV    \\
(8) & Mend & Cor & 1   & No  & Cor & 1   & No  & Cor & 1    & No  & High & Low & High & & 1,000 & CV    \\
(9) & Mend/Platt & Mis & 0.5 & No & Mis & 0.5 & No  & No  &      & No  & High & Low & High & & 1,000 & CV \\
(1)  & Mend & Mis & 0.5 & No  & Mis & 0.5 & No  & No  &      & No  & High & High & Low &  & 10,000 & CV   \\
(2)    & GB   & No  &     & Yes & No  &     & Yes & No  &      & No  & High & High & Low & 50 & 10,000 & CV \\
(3)    & Mend/GB & Mis & 0.5 & Yes & Mis & 0.5 & Yes & No  &      & No  & High & High & Low & 50 & 10,000 & CV  \\
(4)    & Mend & Mis & 0.5 & No  & Mis & 0.5 & No  & Cor & 1    & No  & High & High & Low &  & 10,000 & CV   \\
(5)    & GB   & No  &     & Yes & No  &     & Yes & No  &      & Yes & High & High & Low & 50 & 10,000 & CV \\
(6)    & Mend/GB & Mis & 0.5 & Yes & Mis & 0.5 & Yes & Cor & 1    & Yes & High & High & Low & 50 & 10,000 & CV \\
(7)    & Mend & Cor & 1   & No  & Cor & 1   & No  & No  &      & No  & High & High & Low & & 10,000 & CV    \\
(8)    & Mend & Cor & 1   & No  & Cor & 1   & No  & Cor & 1    & No  & High & High & Low &  & 10,000 & CV   \\
(9) & Mend/Platt & Mis & 0.5 & No & Mis & 0.5 & No  & No  &      & No  & High & High & Low & & 10,000 & CV \\
(10) & Mend/IR & Mis & 0.5 & No  & Mis & 0.5 & No  & No  &      & No  & High & High & Low & & 10,000 & CV \\
(11)    & Mend/GB & Mis  &  0.5   & Yes & Mis  &   0.5  & Yes & No  &      & Yes & Low & Low & High  & 25 & 10,000 & CV \\
 (12)   & Mend/GB & Mis  &  0.5   & Yes & Mis  &  0.5   & Yes & No  &     & Yes & Low & Low & High  & 100 & 10,000 & CV \\
(13)    & GB   & No  &     & Yes & No  &     & Yes & No  &      & Yes & Low & Low & High  & 25 & 10,000 & CV \\
 (14)   & GB   & No  &     & Yes & No  &     & Yes & No  &      & Yes & Low & Low & High  & 100 & 10,000 & CV \\
 (11)   & Mend/GB & Mis  &  0.5   & Yes & Mis  &   0.5  & Yes & No  &      & Yes & High & High & High & 25 & 10,000 & CV \\
 (12)   & Mend/GB & Mis  &  0.5   & Yes & Mis  &  0.5   & Yes & No  &      & Yes & High & High & High & 100 & 10,000 & CV \\
 (13)   & GB   & No  &     & Yes & No  &     & Yes & No  &      & Yes & High & High & High & 25 & 10,000 & CV \\
 (14)   & GB   & No  &     & Yes & No  &     & Yes & No  &      & Yes & High & High & High & 100 & 10,000 & CV \\
 (15)   & Mend & Mis & 0.5 & No  & Mis & 0.5 & No  & Mis & 0.25 & No  & Low & Low & High  &  & 10,000 & CV   \\
 (16)   & Mend & Mis & 0.5 & No  & Mis & 0.5 & No  & Mis & 0.5  & No  & Low & Low & High  &  & 10,000 & CV   \\
 (17)   & Mend & Mis & 0.5 & No  & Mis & 0.5 & No  & Mis & 2    & No  & Low & Low & High  & & 10,000 & CV    \\
 (18)   & Mend & Mis & 0.5 & No  & Mis & 0.5 & No  & Mis & 4    & No  & Low & Low & High  &  & 10,000 & CV   \\
 (15)   & Mend & Mis & 0.5 & No  & Mis & 0.5 & No  & Mis & 0.25 & No  & High & High & High &  & 10,000 & CV   \\
 (16)   & Mend & Mis & 0.5 & No  & Mis & 0.5 & No  & Mis & 0.5  & No  & High & High & High & & 10,000 & CV    \\
 (17)   & Mend & Mis & 0.5 & No  & Mis & 0.5 & No  & Mis & 2    & No  & High & High & High &  & 10,000 & CV   \\
 (18)   & Mend & Mis & 0.5 & No  & Mis & 0.5 & No  & Mis & 4    & No  & High & High & High &  & 10,000 & CV       \\
\bottomrule
\end{tabular}
\end{center}
\end{scriptsize}
\end{table}

\end{landscape}

\begin{table}
\caption{Summary of the simulated data sets for the low- and high-penetrance scenarios. Each entry is the average over the 100 simulated data sets.}
\label{stab:simfams}

\begin{center}
\begin{tabular}{lll}
  \toprule
  & Low-penetrance & High-penetrance \\
  \midrule
  \textbf{General} & & \\
  Number of families & 10,000 & 10,000 \\
  Average family size & 32.00 & 32.01 \\
  Number of male participants & 4991.64 & 4993.87 \\
  \addlinespace
  \textbf{Participant Cancer History} & & \\
  \textit{Number of participants with:} & & \\
  Colorectal cancer (average age) & 38.47 (38.92) & 70.15 (39.05) \\
  Endometrial cancer (average age) & 9.91 (41.32) & 15.23 (41.20) \\
  Gastric cancer (average age) & 25.63 (38.38) & 24.99 (38.32) \\
  \addlinespace
  {\bf Family Member Cancer History} & & \\
  \textit{Number of families with at least} && \\
  \textit{one non-participant with:} & & \\
  Colorectal cancer & 3458.16 & 4264.87 \\
  Endometrial cancer & 1722.32 & 2262.97 \\
  Gastric cancer & 3537.27 & 3539.41 \\
  \addlinespace
  \textbf{Participant MMR Mutations} & & \\
  \textit{Number of participants with mutations of:} & & \\
  \textit{MLH1} & 198.67 & 200.11 \\
  \textit{MSH2} & 200.10 & 199.11 \\
  \textit{MSH6} & 202.37 & 200.71 \\
   \bottomrule
\end{tabular}
\end{center}
\end{table}

\begin{table}
\caption{Calibration intercepts and slopes for the simulated data. For each metric, we provide the mean across the 100 simulated data sets (each data set with 10,000 families), the 95\% confidence interval (CI), and the improvement percentage (IP), or the percentage of simulated data sets where the model outperformed model 1. The Mendelian model is MMRpro. The oracle Mendelian models use the data-generating CRC and EC penetrances. All gradient boosting models are run with 50 iterations.}
\label{stab:cal}

\begin{center}
\begin{tabular}{lcccccc}
 \toprule
\multirow{2}{*}[-0.5em]{Model} & \multicolumn{3}{c}{Intercept} & \multicolumn{3}{c}{Slope} \\ 
\cmidrule(lr){2-4}\cmidrule(lr){5-7}
& Mean & CI & IP & Mean & CI & IP \\
  \midrule
    \multicolumn{7}{l}{\textbf{Low-penetrance data-generating CRC and EC}} \\
    \multicolumn{7}{l}{\textit{Without GC, using misspecified CRC and EC penetrances}} \\
    \ta & $-0.186$ & $(-0.275, -0.122)$ & - & 1.230 & (1.138, 1.316) & - \\ 
  \tb & $-0.049$ & $(-0.064, -0.028)$ & 100 & 1.015 & (0.991, 1.041) & 100 \\ 
  \tc & 0.000 & $(-0.016, 0.024)$ & 100 & 1.016 & (0.973, 1.052) & 100 \\ 
\addlinespace
    \multicolumn{7}{l}{\textit{With GC, using misspecified CRC and EC penetrances}} \\
    \td & $-0.176$ & $(-0.267, -0.110)$ & 80 & 1.129 & (1.053, 1.206) & 99 \\ 
  \te & $-0.049$ & $(-0.065, -0.031)$ & 100 & 1.057 & (1.032, 1.083) & 100 \\ 
  \tf & 0.000 & $(-0.017, 0.024)$ & 100 & 1.001 & (0.961, 1.038) & 100 \\ 
    \addlinespace
    \multicolumn{7}{l}{\textit{Oracle Mendelian models, using correctly-specified CRC and EC penetrances}} \\
    \tg & $-0.001$ & $(-0.095, 0.070)$ & 99 & 1.002 & (0.926, 1.073) & 100 \\ 
  \thh & 0.000 & $(-0.093, 0.069)$ & 99 & 1.008 & (0.937, 1.081) & 100 \\ 
    \addlinespace
    \multicolumn{7}{l}{\textit{Other updating methods on model 1}} \\
    \ti & 0.000 & $(-0.018, 0.025)$ & 100 & 1.003 & (0.989, 1.023) & 100 \\ 
  \tj & 0.133 & (0.109, 0.164) & 94 & 0.966 & (0.945, 0.983) & 100 \\ 
    \midrule
    \multicolumn{7}{l}{\textbf{Low-penetrance data-generating CRC and EC}} \\
    \multicolumn{7}{l}{\textit{Without GC, using misspecified CRC and EC penetrances}} \\
    \ta & $-0.426$ & $(-0.503, -0.333)$ & - & 1.256 & (1.198, 1.337) & - \\ 
  \tb & $-0.057$ & $(-0.075, -0.038)$ & 100 & 1.070 & (1.051, 1.089) & 100 \\ 
  \tc & $-0.002$ & $(-0.019, 0.018)$ & 100 & 1.052 & (1.017, 1.086) & 100 \\ 
    \addlinespace
    \multicolumn{7}{l}{\textit{With GC, using misspecified CRC and EC penetrances}} \\
    \td & $-0.387$ & $(-0.468, -0.298)$ & 100 & 1.228 & (1.161, 1.295) & 86 \\ 
  \te & $-0.057$ & $(-0.074, -0.037)$ & 100 & 1.086 & (1.065, 1.106) & 100 \\ 
  \tf & $-0.001$ & $(-0.017, 0.019)$ & 100 & 1.048 & (1.018, 1.089) & 100 \\
    \addlinespace
    \multicolumn{7}{l}{\textit{Oracle Mendelian models, using correctly-specified CRC and EC penetrances}} \\
    \tg & 0.004 & $(-0.083, 0.111)$ & 100 & 1.007 & (0.960, 1.068) & 100 \\ 
  \thh & 0.003 & $(-0.096, 0.103)$ & 100 & 1.010 & (0.956, 1.064) & 100 \\ 
    \addlinespace
    \multicolumn{7}{l}{\textit{Other updating methods on model 1}} \\
    \ti & $-0.001$ & $(-0.020, 0.019)$ & 100 & 1.001 & (0.993, 1.011) & 100 \\ 
  \tj & 0.155 & (0.131, 0.182) & 100 & 0.950 & (0.930, 0.970) & 100 \\ 
   \bottomrule
 \end{tabular}
 \end{center}
\end{table}

\begin{table}
\caption{Sensitivity analysis of the number of iterations in gradient boosting for the simulated data. The top and bottom represent data generated from low (scaled MMRpro) and high (MMRpro) colorectal and endometrial cancer penetrances. All models use information from colorectal, endometrial, and gastric cancers as features. For each metric, we provide the mean across the 100 simulated data sets (each data set with 10,000 families), the 95\% confidence interval (CI), and the improvement percentage (IP), or the percentage of simulated data sets where the model outperformed model 1.}
\label{stab:sim_sens}

\begin{center}
\begin{scriptsize}
\begin{tabular}{lllcccccccccc}
  \toprule
\multirow{2}{*}{Num.} & With & \multirow{2}{*}{Iter.$^*$} & \multicolumn{3}{c}{O/E} & \multicolumn{3}{c}{AUC} & \multicolumn{3}{c}{rBS} \\
\cmidrule(lr){4-6}  \cmidrule(lr){7-9} \cmidrule(lr){10-12}
& Mend.$^\dagger$ & & Mean & CI & IP & Mean & CI & IP & Mean & CI & IP \\
\midrule
\multicolumn{11}{l}{\textbf{Low-penetrance data-generating CRC and EC}} \\
(11) & Yes & 25 & 0.990 & (0.977, 1.007) & 100 & 0.837 & (0.818, 0.852) & 100 & 0.215 & (0.208, 0.221) & 100 \\ 
(6) &  Yes & 50 & 1.003 & (0.990, 1.022) & 100 & 0.836 & (0.818, 0.850) & 100 & 0.215 & (0.208, 0.222) & 100 \\ 
(12) & Yes & 100 & 1.004 & (0.992, 1.023) & 100 & 0.833 & (0.816, 0.848) & 100 & 0.216 & (0.209, 0.223) & 100 \\ 
(13) & No & 25 & 0.630 & (0.607, 0.647) & 0 & 0.793 & (0.775, 0.808) & 78 & 0.227 & (0.220, 0.233) & 0 \\ 
(5) & No & 50 & 0.962 & (0.949, 0.977) & 100 & 0.797 & (0.778, 0.812) & 85 & 0.224 & (0.217, 0.230) & 0 \\ 
(14) & No & 100 & 1.005 & (0.992, 1.025) & 100 & 0.795 & (0.777, 0.812) & 85 & 0.224 & (0.217, 0.230) & 0 \\
  \midrule
  \multicolumn{11}{l}{\textbf{High-penetrance data-generating CRC and EC}} \\
(11) & Yes & 25 & 0.976 & (0.964, 0.988) & 100 & 0.923 & (0.914, 0.932) & 100 & 0.195 & (0.188, 0.202) & 100 \\ 
(6) & Yes & 50 & 1.002 & (0.991, 1.015) & 100 & 0.922 & (0.913, 0.931) & 100 & 0.196 & (0.189, 0.202) & 73 \\ 
(12) & Yes & 100 & 1.005 & (0.993, 1.016) & 100 & 0.921 & (0.912, 0.930) & 98 & 0.197 & (0.190, 0.203) & 17 \\ 
(13) & No & 25 & 0.629 & (0.610, 0.648) & 0 & 0.877 & (0.864, 0.888) & 0 & 0.218 & (0.211, 0.224) & 0 \\ 
(5) & No & 50 & 0.959 & (0.947, 0.973) & 100 & 0.882 & (0.869, 0.892) & 0 & 0.214 & (0.207, 0.221) & 0 \\ 
(14) & No & 100 & 1.006 & (0.993, 1.021) & 100 & 0.882 & (0.868, 0.892) & 0 & 0.215 & (0.207, 0.221) & 0 \\ 
   \bottomrule
\end{tabular}
\end{scriptsize}
\end{center}

\begin{flushleft}
$^\dagger$ Initializing boosting with Mendelian predictions without gastric cancer (model 1) \\
$^*$ Number of iterations in gradient boosting ($M$ from Section \ref{gb})
\end{flushleft}
\end{table}

\begin{table}
\caption{Analysis of Mendelian model performance on the simulated data when incorporating GC using misspecified penetrances. The exponent is applied to the gastric cancer survival function. The top and bottom represent low (scaled MMRpro) and high (MMRpro) colorectal and endometrial cancer penetrances. For all models, the colorectal and endometrial penetrances are misspecified (different from data-generating) by taking the square root of the data-generating survival functions and converting back to penetrances. For each metric, we provide the mean across the 100 simulated data sets (each data set with 10,000 families), the 95\% confidence interval (CI), and the improvement percentage (IP), or the percentage of simulated data sets where the model outperformed model 1.}
\label{stab:sim_mis}

\begin{center}
\begin{scriptsize}
\begin{tabular}{llccccccccc}
  \toprule
\multirow{2}{*}[-0.5em]{Num.} & \multirow{2}{*}[-0.5em]{Exponent$^\dagger$} & \multicolumn{3}{c}{O/E} & \multicolumn{3}{c}{AUC} & \multicolumn{3}{c}{rBS} \\
\cmidrule(lr){3-5}  \cmidrule(lr){6-8} \cmidrule(lr){9-11}
& & Mean & CI & IP & Mean & CI & IP & Mean & CI & IP \\
\midrule
  \multicolumn{10}{l}{\textbf{Low-penetrance data-generating CRC and EC}} \\
(15) & 0.25 & 0.799 & (0.742, 0.843) & 0 & 0.842 & (0.822, 0.858) & 100 & 0.213 & (0.207, 0.219) & 100 \\ 
(16) & 0.5 & 0.798 & (0.741, 0.841) & 0 & 0.849 & (0.830, 0.864) & 100 & 0.212 & (0.205, 0.217) & 100 \\ 
(17) & 2 & 1.104 & (1.030, 1.165) & 73 & 0.840 & (0.823, 0.858) & 100 & 0.213 & (0.206, 0.219) & 100 \\ 
(18) & 4 & 1.688 & (1.583, 1.796) & 0 & 0.827 & (0.809, 0.845) & 100 & 0.217 & (0.209, 0.223) & 72 \\ 
  \midrule
  \multicolumn{10}{l}{\textbf{High-penetrance data-generating CRC and EC}} \\
 (15) & 0.25 & 0.714 & (0.677, 0.763) & 0 & 0.928 & (0.918, 0.937) & 100 & 0.194 & (0.188, 0.201) & 100 \\ 
 (16) & 0.5 & 0.722 & (0.684, 0.770) & 0 & 0.930 & (0.920, 0.938) & 100 & 0.193 & (0.187, 0.200) & 100 \\ 
 (17) & 2 & 0.902 & (0.853, 0.959) & 100 & 0.926 & (0.916, 0.935) & 100 & 0.193 & (0.186, 0.200) & 99 \\ 
 (18) & 4 & 1.197 & (1.128, 1.276) & 86 & 0.917 & (0.907, 0.928) & 55 & 0.196 & (0.189, 0.203) & 63 \\ 
   \bottomrule
\end{tabular}
\end{scriptsize}
\end{center}

\begin{flushleft}
$^\dag$ Exponent applied to the gastric cancer survival function
\end{flushleft}
\end{table}

\begin{table}
\caption{Performance measures for the simulated data with small sample size. For each metric, we provide the mean across the 100 simulated data sets (each data set with 1,000 families), the 95\% confidence interval (CI), and the improvement percentage (IP), or the percentage of simulated data sets where the model outperformed model 1. The Mendelian model is MMRpro. The oracle Mendelian models use the data-generating CRC and EC penetrances. All gradient boosting models are run with 50 iterations.}
\label{stab:sss}

\begin{center}
\begin{scriptsize}
\begin{tabular}{lccccccccc}
  \toprule
\multirow{2}{*}[-0.5em]{Model} & \multicolumn{3}{c}{O/E} & \multicolumn{3}{c}{AUC} & \multicolumn{3}{c}{rBS} \\ 
\cmidrule(lr){2-4}\cmidrule(lr){5-7}\cmidrule(lr){8-10}
& Mean & CI & IP & Mean & CI & IP & Mean & CI & IP \\
  \midrule
  \multicolumn{10}{l}{\textbf{Low-penetrance data-generating CRC and EC}} \\
  \multicolumn{10}{l}{\textit{Without GC, using misspecified CRC and EC penetrances}} \\
  \ta & 0.855 & (0.681, 1.046) & - & 0.783 & (0.715, 0.852) & - & 0.218 & (0.195, 0.242) & - \\ 
  \tb & 0.982 & (0.929, 1.036) & 96 & 0.695 & (0.616, 0.766) & 0 & 0.231 & (0.206, 0.256) & 0 \\ 
  \tc & 1.052 & (0.999, 1.120) & 82 & 0.761 & (0.689, 0.829) & 7 & 0.222 & (0.198, 0.246) & 1 \\ 
\addlinespace
    \multicolumn{10}{l}{\textit{With GC, using misspecified CRC and EC penetrances}} \\
   \td & 0.868 & (0.698, 1.037) & 70 & 0.847 & (0.794, 0.899) & 99 & 0.211 & (0.190, 0.233) & 94 \\ 
  \te & 0.986 & (0.933, 1.042) & 95 & 0.764 & (0.692, 0.824) & 31 & 0.228 & (0.205, 0.251) & 2 \\ 
  \tf & 1.057 & (1.005, 1.131) & 83 & 0.810 & (0.746, 0.875) & 87 & 0.220 & (0.196, 0.244) & 18 \\
\addlinespace
    \multicolumn{10}{l}{\textit{Oracle Mendelian models, using correctly-specified CRC and EC penetrances}} \\
    \tg & 1.001 & (0.803, 1.230) & 72 & 0.783 & (0.710, 0.846) & 47 & 0.217 & (0.195, 0.241) & 79 \\ 
  \thh & 0.998 & (0.800, 1.198) & 70 & 0.849 & (0.793, 0.898) & 99 & 0.210 & (0.188, 0.234) & 97 \\ 
    \addlinespace
    \multicolumn{10}{l}{\textit{Other updating methods on model 1}} \\
    \ti & 1.028 & (0.976, 1.091) & 90 & 0.783 & (0.715, 0.852) & 0 & 0.221 & (0.196, 0.246) & 8 \\ 
  \tj & 1.308 & (1.195, 1.438) & 9 & 0.783 & (0.714, 0.852) & 5 & 0.221 & (0.197, 0.245) & 6 \\ 

   \midrule
   \multicolumn{10}{l}{\textbf{High-penetrance data-generating CRC and EC}} \\
    \multicolumn{10}{l}{\textit{Without GC, using misspecified CRC and EC penetrances}} \\
   \ta & 0.729 & (0.574, 0.886) & - & 0.916 & (0.864, 0.945) & - & 0.195 & (0.176, 0.213) & - \\ 
  \tb & 0.977 & (0.924, 1.020) & 100 & 0.852 & (0.782, 0.898) & 0 & 0.219 & (0.196, 0.241) & 0 \\ 
  \tc & 1.051 & (1.001, 1.108) & 100 & 0.905 & (0.850, 0.935) & 1 & 0.200 & (0.178, 0.222) & 7 \\ 
\addlinespace
  \multicolumn{10}{l}{\textit{With GC, using misspecified CRC and EC penetrances}} \\
  \td & 0.759 & (0.602, 0.918) & 99 & 0.931 & (0.897, 0.957) & 94 & 0.191 & (0.174, 0.208) & 93 \\ 
  \te & 0.983 & (0.935, 1.027) & 100 & 0.863 & (0.804, 0.904) & 0 & 0.219 & (0.196, 0.240) & 0 \\ 
  \tf & 1.060 & (1.008, 1.121) & 100 & 0.912 & (0.870, 0.942) & 31 & 0.200 & (0.178, 0.222) & 7 \\ 
\addlinespace
  \multicolumn{10}{l}{\textit{Oracle Mendelian models, using correctly-specified CRC and EC penetrances}} \\
  \tg & 0.987 & (0.782, 1.207) & 93 & 0.917 & (0.866, 0.946) & 70 & 0.192 & (0.171, 0.212) & 94 \\ 
  \thh & 0.987 & (0.793, 1.207) & 94 & 0.932 & (0.897, 0.956) & 99 & 0.188 & (0.169, 0.206) & 97 \\ 
\addlinespace
    \multicolumn{10}{l}{\textit{Other updating methods on model 1}} \\
    \ti & 1.027 & (0.987, 1.066) & 100 & 0.916 & (0.864, 0.945) & 0 & 0.196 & (0.173, 0.218) & 24 \\ 
  \tj & 1.239 & (1.124, 1.367) & 64 & 0.916 & (0.864, 0.945) & 0 & 0.197 & (0.174, 0.217) & 13 \\ 
   \bottomrule
\end{tabular}
\end{scriptsize}
\end{center}
\end{table}

\begin{table}
\caption{Performance measures for the simulated data, using bootstrap validation instead of Monte Carlo cross-validation. For each metric, we provide the mean across the 100 simulated data sets (each data set with 10,000 families), the 95\% confidence interval (CI), and the improvement percentage (IP), or the percentage of simulated data sets where the model outperformed model 1. The Mendelian model is MMRpro. The oracle Mendelian models use the data-generating CRC and EC penetrances. All gradient boosting models are run with 50 iterations.}
\label{stab:boot}

\begin{center}
\begin{scriptsize}
\begin{tabular}{lccccccccc}
 \toprule
\multirow{2}{*}[-0.5em]{Model} & \multicolumn{3}{c}{O/E} & \multicolumn{3}{c}{AUC} & \multicolumn{3}{c}{rBS} \\ 
\cmidrule(lr){2-4}\cmidrule(lr){5-7}\cmidrule(lr){8-10}
& Mean & CI & IP & Mean & CI & IP & Mean & CI & IP \\
  \midrule
    \multicolumn{10}{l}{\textbf{Low-penetrance data-generating CRC and EC}} \\
    \multicolumn{10}{l}{\textit{Without GC, using misspecified CRC and EC penetrances}} \\
    \ta & 0.855 & (0.790, 0.903) & - & 0.785 & (0.762, 0.805) & - & 0.218 & (0.211, 0.224) & - \\ 
  \tb & 0.960 & (0.945, 0.978) & 100 & 0.735 & (0.712, 0.755) & 0 & 0.228 & (0.219, 0.234) & 0 \\ 
  \tc & 1.001 & (0.983, 1.023) & 100 & 0.784 & (0.760, 0.804) & 36 & 0.218 & (0.211, 0.225) & 29 \\
  \addlinespace
    \multicolumn{10}{l}{\textit{With GC, using misspecified CRC and EC penetrances}} \\
    \td & 0.871 & (0.811, 0.918) & 97 & 0.848 & (0.831, 0.864) & 100 & 0.211 & (0.205, 0.217) & 100 \\ 
  \te & 0.959 & (0.944, 0.974) & 100 & 0.800 & (0.782, 0.816) & 92 & 0.223 & (0.215, 0.229) & 0 \\ 
  \tf & 1.001 & (0.985, 1.020) & 100 & 0.838 & (0.821, 0.853) & 100 & 0.215 & (0.208, 0.221) & 100 \\ 
    \addlinespace
    \multicolumn{10}{l}{\textit{Oracle Mendelian models, using correctly-specified CRC and EC penetrances}} \\
    \tg & 1.000 & (0.925, 1.057) & 99 & 0.787 & (0.763, 0.805) & 79 & 0.217 & (0.210, 0.224) & 99 \\ 
  \thh & 1.000 & (0.933, 1.055) & 99 & 0.851 & (0.834, 0.867) & 100 & 0.211 & (0.204, 0.217) & 100 \\ 
   \addlinespace
    \multicolumn{10}{l}{\textit{Other updating methods on model 1}} \\
    \ti & 1.001 & (0.993, 1.008) & 100 & 0.785 & (0.762, 0.805) & 0 & 0.220 & (0.212, 0.227) & 0 \\ 
  \tj & 1.084 & (1.049, 1.124) & 96 & 0.785 & (0.762, 0.804) & 17 & 0.217 & (0.210, 0.224) & 100 \\ 
   \midrule
   \multicolumn{10}{l}{\textbf{High-penetrance data-generating CRC and EC}} \\
    \multicolumn{10}{l}{\textit{Without GC, using misspecified CRC and EC penetrances}} \\
    \ta & 0.742 & (0.702, 0.798) & - & 0.917 & (0.906, 0.927) & - & 0.196 & (0.190, 0.202) & - \\ 
  \tb & 0.958 & (0.941, 0.974) & 100 & 0.871 & (0.856, 0.882) & 0 & 0.215 & (0.207, 0.222) & 0 \\ 
  \tc & 0.999 & (0.983, 1.015) & 100 & 0.915 & (0.905, 0.925) & 13 & 0.196 & (0.189, 0.202) & 82 \\ 
    \addlinespace
    \multicolumn{10}{l}{\textit{With GC, using misspecified CRC and EC penetrances}} \\
    \td & 0.771 & (0.729, 0.824) & 100 & 0.930 & (0.920, 0.938) & 100 & 0.193 & (0.186, 0.199) & 100 \\ 
  \te & 0.959 & (0.941, 0.974) & 100 & 0.884 & (0.872, 0.894) & 0 & 0.214 & (0.206, 0.221) & 0 \\ 
  \tf & 1.001 & (0.983, 1.020) & 100 & 0.924 & (0.915, 0.933) & 100 & 0.195 & (0.188, 0.201) & 99 \\ 
    \addlinespace
    \multicolumn{10}{l}{\textit{Oracle Mendelian models, using correctly-specified CRC and EC penetrances}} \\
    \tg & 1.004 & (0.946, 1.079) & 100 & 0.917 & (0.907, 0.928) & 95 & 0.194 & (0.187, 0.200) & 100 \\ 
  \thh & 1.003 & (0.941, 1.069) & 100 & 0.931 & (0.921, 0.940) & 100 & 0.191 & (0.184, 0.198) & 100 \\ 
    \addlinespace
    \multicolumn{10}{l}{\textit{Other updating methods on model 1}} \\
    \ti & 1.002 & (0.994, 1.008) & 100 & 0.917 & (0.906, 0.927) & 0 & 0.197 & (0.189, 0.204) & 14 \\ 
  \tj & 1.074 & (1.024, 1.114) & 100 & 0.917 & (0.907, 0.927) & 55 & 0.194 & (0.187, 0.200) & 100 \\ 
   \bottomrule
 \end{tabular}
\end{scriptsize}
\end{center}
\end{table}

\begin{table}
\caption{Performance measures for the simulated data where the training data is generated using high-penetrance CRC and EC, and the testing data is generating using penetrances obtained by taking the square root of the data-generating survival functions for the training data. For each metric, we provide the mean across the 100 simulated data sets (each data set with 10,000 families), the 95\% confidence interval (CI), and the improvement percentage (IP), or the percentage of simulated data sets where the model outperformed model 1. The Mendelian model is MMRpro. The top and bottom represent data generated from low and high CRC and EC penetrances, respectively. The oracle Mendelian models use the data-generating CRC and EC penetrances. All gradient boosting models are run with 50 iterations.}
\label{stab:trans}

\begin{center}
\begin{scriptsize}
\textbf{Training and testing data sets generated using different penetrances} \\
\begin{tabular}{lccccccccc}
 \toprule
\multirow{2}{*}[-0.5em]{Model} & \multicolumn{3}{c}{O/E} & \multicolumn{3}{c}{AUC} & \multicolumn{3}{c}{rBS} \\ 
\cmidrule(lr){2-4}\cmidrule(lr){5-7}\cmidrule(lr){8-10}
& Mean & CI & IP & Mean & CI & IP & Mean & CI & IP \\
  \midrule
    \multicolumn{10}{l}{\textit{Without GC, using misspecified CRC and EC penetrances}} \\
    \ta & 1.175 & (1.065, 1.298) & - & 0.843 & (0.819, 0.868) & - & 0.211 & (0.200, 0.222) & - \\ 
  \tb & 1.747 & (1.532, 2.013) & 0 & 0.803 & (0.779, 0.832) & 0 & 0.225 & (0.214, 0.238) & 0 \\ 
  \tc & 1.821 & (1.570, 2.117) & 0 & 0.841 & (0.817, 0.869) & 17 & 0.215 & (0.204, 0.226) & 0 \\ 
    \addlinespace
    \multicolumn{10}{l}{\textit{With GC, using misspecified CRC and EC penetrances}} \\
    \td & 1.374 & (1.244, 1.521) & 0 & 0.865 & (0.843, 0.886) & 100 & 0.209 & (0.199, 0.221) & 86 \\ 
  \te & 1.879 & (1.655, 2.182) & 0 & 0.826 & (0.801, 0.853) & 4 & 0.225 & (0.213, 0.238) & 0 \\ 
  \tf & 1.979 & (1.746, 2.325) & 0 & 0.860 & (0.836, 0.884) & 98 & 0.215 & (0.204, 0.227) & 0 \\
    \addlinespace
    \multicolumn{10}{l}{\textit{Oracle Mendelian models, using correctly-specified CRC and EC penetrances}} \\
    \tg & 1.805 & (1.616, 1.988) & 0 & 0.843 & (0.818, 0.868) & 32 & 0.214 & (0.202, 0.225) & 0 \\ 
  \thh & 1.945 & (1.743, 2.154) & 0 & 0.864 & (0.845, 0.887) & 100 & 0.212 & (0.202, 0.225) & 4 \\ 
    \addlinespace
    \multicolumn{10}{l}{\textit{Other updating methods on model 1}} \\
    \ti & 1.693 & (1.500, 1.958) & 0 & 0.843 & (0.819, 0.868) & 0 & 0.217 & (0.205, 0.228) & 0 \\ 
  \tj & 2.063 & (1.788, 2.411) & 0 & 0.843 & (0.819, 0.868) & 16 & 0.215 & (0.204, 0.227) & 0 \\ 
   \bottomrule
 \end{tabular}
 \end{scriptsize}
 \end{center}
\end{table}

\begin{table}
\caption{Performance measures for the simulated data, using low data-generating GC penetrances, with high data-generating CRC and EC penetrances, and a sample size of 1000 families. Here the carrier GC penetrances are 2 times the noncarrier GC penetrances. For each metric, we provide the mean across the 100 simulated data sets (each data set with 10,000 families), the 95\% confidence interval (CI), and the improvement percentage (IP), or the percentage of simulated data sets where the model outperformed model 1. The Mendelian model is MMRpro. The oracle Mendelian models use the data-generating CRC and EC penetrances. All gradient boosting models are run with 50 iterations.}
\label{stab:lgc}

\begin{center}
\begin{scriptsize}
\begin{tabular}{lccccccccc}
 \toprule
\multirow{2}{*}[-0.5em]{Model} & \multicolumn{3}{c}{O/E} & \multicolumn{3}{c}{AUC} & \multicolumn{3}{c}{rBS} \\ 
\cmidrule(lr){2-4}\cmidrule(lr){5-7}\cmidrule(lr){8-10}
& Mean & CI & IP & Mean & CI & IP & Mean & CI & IP \\
  \midrule
    \multicolumn{10}{l}{\textit{Without GC, using misspecified CRC and EC penetrances}} \\
    \ta & 0.735 & (0.600, 0.879) & 0 & 0.914 & (0.869, 0.949) & 0 & 0.195 & (0.176, 0.216) & 0 \\ 
  \tb & 0.974 & (0.923, 1.021) & 100 & 0.846 & (0.798, 0.890) & 0 & 0.220 & (0.197, 0.241) & 0 \\ 
 \tc & 1.049 & (1.002, 1.110) & 99 & 0.903 & (0.857, 0.943) & 3 & 0.201 & (0.180, 0.222) & 2 \\ 
    \addlinespace
    \multicolumn{10}{l}{\textit{With GC, using misspecified CRC and EC penetrances}} \\
   \td & 0.735 & (0.599, 0.879) & 48 & 0.914 & (0.871, 0.951) & 35 & 0.195 & (0.176, 0.216) & 47 \\ 
  \te & 0.976 & (0.925, 1.023) & 100 & 0.846 & (0.798, 0.892) & 0 & 0.220 & (0.198, 0.241) & 0 \\ 
  \tf & 1.050 & (1.008, 1.115) & 100 & 0.903 & (0.857, 0.943) & 3 & 0.201 & (0.180, 0.222) & 2 \\ 
    \addlinespace
    \multicolumn{10}{l}{\textit{Oracle Mendelian models, using correctly-specified CRC and EC penetrances}} \\
    \tg & 0.998 & (0.819, 1.211) & 92 & 0.915 & (0.872, 0.950) & 59 & 0.193 & (0.173, 0.216) & 94 \\ 
  \thh & 0.998 & (0.818, 1.211) & 92 & 0.915 & (0.872, 0.951) & 62 & 0.193 & (0.173, 0.216) & 94 \\ 
    \addlinespace
    \multicolumn{10}{l}{\textit{Other updating methods on model 1}} \\
  \ti & 1.024 & (0.981, 1.073) & 100 & 0.914 & (0.869, 0.949) & 0 & 0.197 & (0.176, 0.223) & 31 \\ 
  \tj & 1.237 & (1.127, 1.359) & 61 & 0.914 & (0.869, 0.949) & 0 & 0.198 & (0.175, 0.220) & 11 \\ 
   \bottomrule
 \end{tabular}
 \end{scriptsize}
 \end{center}
\end{table}

\begin{table}
\caption{Calibration intercepts and slopes for the USC-Stanford data, with 95\% confidence intervals. For each metric, we provide the mean across the 1000 Monte Carlo cross-validation replicates, the 95\% confidence interval (CI), and the improvement percentage (IP), or the percentage of Monte Carlo cross-validation replicates where the model outperformed model 1. The Mendelian model is MMRpro. All gradient boosting models are run with 50 iterations.}
\label{stab:cal_usc}

\begin{center}
\begin{tabular}{lcccccc}
 \toprule
\multirow{2}{*}[-0.5em]{Model} & \multicolumn{3}{c}{Intercept} & \multicolumn{3}{c}{Slope} \\ 
\cmidrule(lr){2-4}\cmidrule(lr){5-7}
& Mean & CI & IP & Mean & CI & IP \\
  \midrule
    \ta & $-0.302$ & $(-1.124, 0.374)$ & 0 & 0.408 & (0.314, 0.504) & 0 \\ 
  \tb & $-0.282$ & $(-1.005, 0.376)$ & 53 & 0.842 & (0.326, 1.615) & 93 \\ 
  \tc & $-0.149$ & $(-1.414, 1.147)$ & 35 & 0.448 & (0.310, 0.608) & 89 \\ 
    \addlinespace
    \multicolumn{7}{l}{\textit{With GC, using misspecified CRC and EC penetrances}} \\
    \td & $-0.562$ & $(-1.407, 0.164)$ & 12 & 0.381 & (0.283, 0.481) & 2 \\ 
  \te & $-0.285$ & $(-1.015, 0.383)$ & 52 & 0.818 & (0.319, 1.518) & 93 \\ 
  \tf & $-0.162$ & $(-1.412, 1.141)$ & 32 & 0.447 & (0.308, 0.602) & 90 \\
    \addlinespace
    \multicolumn{7}{l}{\textit{Other updating methods on model 1}} \\
    \ti & $-0.030$ & $(-0.884, 0.795)$ & 62 & 1.263 & (0.483, 1.950) & 90 \\ 
  \tj & 0.408 & $(-0.771, 2.617)$ & 47 & 0.984 & (0.397, 2.348) & 89 \\ 
  \bottomrule
 \end{tabular}
 \end{center}
\end{table}

\begin{figure}
    \label{sfig:cal_usc}
    \centering
    \includegraphics[width=\linewidth]{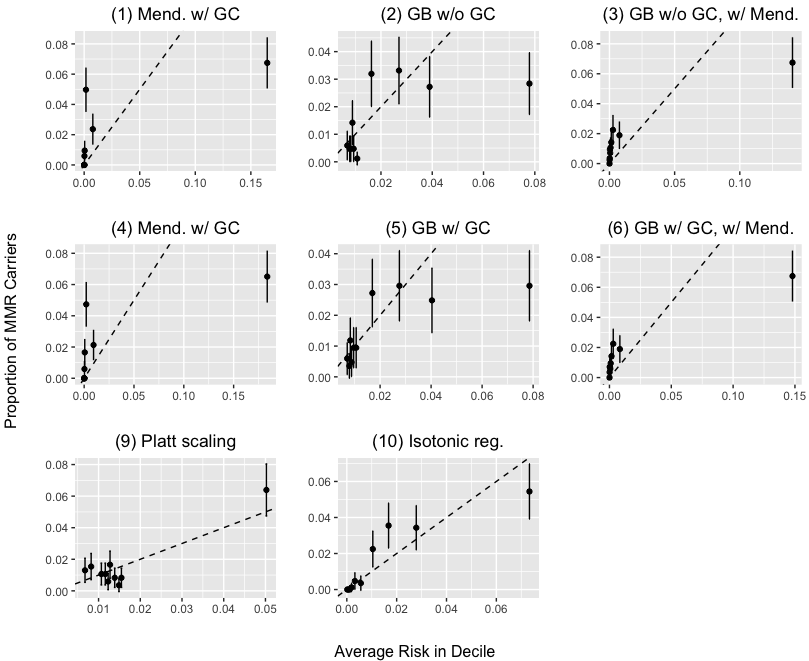}
    \caption{Calibration plots for the USC-Stanford data, using 10 Monte Carlo cross-validation replicates. Each point represents a risk decile, and the error bars are the binomial confidence intervals for the observed proportion of carriers. Each family may appear in the plot multiple times if they were in the testing set multiple times.}
\end{figure}

\begin{table}[hbt!]
\caption{Sensitivity analysis of the number of iterations in gradient boosting for the USC-Stanford data. All models use information from colorectal, endometrial, and gastric cancers as features. For each metric, we provide the mean across the 1000 Monte Carlo cross-validation replicates, the 95\% confidence interval (CI), and the improvement percentage (IP), or the percentage of Monte Carlo cross-validation replicates where the model outperformed model 1.}
\label{stab:usc_sens}

\begin{center}
\begin{footnotesize}
\begin{tabular}{llcccccccccc}
  \toprule
With & \multirow{2}{*}{Iter.$^*$} & \multicolumn{3}{c}{O/E} & \multicolumn{3}{c}{AUC} & \multicolumn{3}{c}{rBS} \\
\cmidrule(lr){3-5}  \cmidrule(lr){6-8} \cmidrule(lr){9-11}
Mend.$^\dagger$ & & Mean & CI & IP & Mean & CI & IP & Mean & CI & IP \\
\midrule
Yes & 25 & 0.949 & (0.418, 1.818) & 28 & 0.839 & (0.747, 0.902) & 24 & 0.130 & (0.112, 0.147) & 72 \\ 
  Yes & 50 & 0.985 & (0.419, 1.941) & 27 & 0.815 & (0.685, 0.895) & 14 & 0.129 & (0.111, 0.148) & 75 \\ 
  Yes & 100 & 1.006 & (0.430, 2.116) & 30 & 0.790 & (0.633, 0.882) & 8 & 0.129 & (0.109, 0.147) & 78 \\ 
  No & 25 & 0.279 & (0.163, 0.411) & 0 & 0.691 & (0.553, 0.803) & 0 & 0.131 & (0.114, 0.148) & 56 \\ 
  No & 50 & 0.807 & (0.377, 1.441) & 21 & 0.716 & (0.620, 0.809) & 0 & 0.125 & (0.104, 0.144) & 86 \\ 
  No & 100 & 1.058 & (0.458, 2.142) & 34 & 0.703 & (0.580, 0.811) & 0 & 0.125 & (0.103, 0.145) & 84 \\ 
  \bottomrule
\end{tabular}
\end{footnotesize}
\end{center}

\begin{flushleft}
$^\dag$ Initializing boosting with the Mendelian predictions without gastric cancer (model 1) \\
$^\ddag$ Number of iterations in gradient boosting ($M$ from Section \ref{gb})
\end{flushleft}

\end{table}

\begin{table}[hbt!]
\caption{Analysis of Mendelian model performance on the USC-Stanford data when incorporating gastric cancer using altered penetrances. The exponent is applied to the gastric cancer survival function. For each metric, we provide the mean across the 1000 Monte Carlo cross-validation replicates, the 95\% confidence interval (CI), and the improvement percentage (IP), or the percentage of Monte Carlo cross-validation replicates where the model outperformed model 1.}
\label{stab:usc_mis}

\begin{center}
\begin{tabular}{lccccccccc}
  \toprule
\multirow{2}{*}[-0.5em]{Exponent$^\dagger$} & \multicolumn{3}{c}{O/E} & \multicolumn{3}{c}{AUC} & \multicolumn{3}{c}{rBS} \\
\cmidrule(lr){2-4}  \cmidrule(lr){5-7} \cmidrule(lr){8-10}
& Mean & CI & IP & Mean & CI & IP & Mean & CI & IP \\
\midrule
0.25 & 0.722 & (0.445, 1.039) & 9 & 0.845 & (0.793, 0.895) & 0 & 0.139 & (0.119, 0.159) & 0 \\ 
  0.5 & 0.735 & (0.452, 1.058) & 10 & 0.845 & (0.793, 0.895) & 0 & 0.139 & (0.118, 0.158) & 0 \\ 
  2 & 0.809 & (0.495, 1.166) & 19 & 0.843 & (0.791, 0.895) & 0 & 0.136 & (0.115, 0.156) & 8 \\ 
  4 & 0.907 & (0.553, 1.319) & 66 & 0.843 & (0.792, 0.895) & 0 & 0.133 & (0.112, 0.153) & 34 \\ 
  \bottomrule
\end{tabular}
\end{center}

\begin{flushleft}
$^\dag$ Exponent applied to the gastric cancer survival function
\end{flushleft}

\end{table}

\end{document}